\documentclass{aa}
\usepackage{graphicx}
\usepackage[varg]{txfonts}
\usepackage[dvipsnames]{xcolor}
\usepackage[normalem]{ulem}

\begin{document}
\newcommand{\be}{\begin{equation}}
\newcommand{\ee}{\end{equation}}
\newcommand{\bq}{\begin{eqnarray}}
\newcommand{\eq}{\end{eqnarray}}

\title{Primordial nucleosynthesis with varying fundamental constants}
\subtitle{Solutions to the Lithium problem and the Deuterium discrepancy}

\author{M. Deal\inst{1,2} \and C. J. A. P. Martins\inst{1,2}}
\institute{Centro de Astrof\'{\i}sica da Universidade do Porto, Rua das Estrelas, 4150-762 Porto, Portugal\\
\email{Morgan.Deal@astro.up.pt, Carlos.Martins@astro.up.pt}
\and
Instituto de Astrof\'{\i}sica e Ci\^encias do Espa\c co, CAUP, Rua das Estrelas, 4150-762 Porto, Portugal}
\date{Submitted \today}

\abstract
{The success of primordial nucleosynthesis has been limited by the long-standing Lithium problem. We use a self-consistent perturbative analysis of the effects of relevant theoretical parameters on primordial nucleosynthesis, including variations of nature's fundamental constants, to explore the problem and its possible solutions, in the context of the latest observations and theoretical modeling. We quantify the amount of depletion needed to solve the Lithium problem, and show that transport processes of chemical elements in stars are able to account for it. Specifically, the combination of atomic diffusion, rotation and penetrative convection allows us to reproduce the lithium surface abundances of Population II stars, starting from the primordial Lithium abundance. We also show that even with this depletion factor there is a preference for a value of the fine-structure constant at this epoch that is larger than the current laboratory one by a few parts per million of relative variation, at the two to three standard deviations level of statistical significance. This preference is driven by the recently noticed discrepancy between the best-fit values for the baryon-to-photon ratio (or equivalently the Deuterium abundance) inferred from cosmic microwave background and primordial nucleosynthesis analyses, and is largely insensitive to the Helium-4 abundance. We thus conclude that the Lithium problem most likely has an astrophysical solution, while the Deuterium discrepancy provides a possible hint of new physics.}


\titlerunning{Primordial Nucleosynthesis with Varying Fundamental Constants}
\authorrunning{Deal \& Martins}
\maketitle

%%%%%%%%%%%%%%%%%%%%%%%%%%%%%%%%%%%%%%%%%%%%%%%%%%%%%%%%%%%%%%%%%%%%%%%%%%
\section{Introduction}
\label{introd}

Big Bang Nucleosynthesis (henceforth BBN) is a cornerstone of the standard particle cosmology paradigm, and a sensitive probe of physics beyond the standard model \citep{Steigman,Iocco,Pitrou}. Nevertheless, its success is limited by the well-known Lithium problem, wherein the theoretically expected abundance of Lithium-7 (given our present knowledge of astrophysics, nuclear and particle physics) exceeds the observed one by a factor of about 3.5 \citep{PDG}. Although there have been many attempts to solve the problem, there is no clear known solution. Further discussion of these solutions can be found in \citet{Fields}, \citet{Mathews} and the BBN section of the latest Particle Data Group (henceforth PDG) review by \citet{PDG}.

Recent progress in experimental measurements of the required nuclear cross-sections has all but excluded the possibility of nuclear physics systematics \citep{Iliadis,Mossa}. On the astrophysics side, some degree of lithium depletion can occur in stars due to the mixing of the outer layers with the hotter interior \citep{sbordone10}, and some authors have suggested that a depletion by factor as large as 1.8 may have occurred \citep{Ryan,Korn}. It has been argued that this scenario might be difficult to reconcile with the existence of extremely iron-poor dwarf stars with lithium abundances very close to the Spite plateau \citep{Aguado}, but there is no consensus on this point. Finally, the Lithium problem could also point to new physics beyond the standard model; one such possibility is the variation of nature's fundamental constants, which is unavoidable in many extensions of the standard particle physics and cosmological models \citep{Damour}---a recent review of the topic is \citet{ROPP}. This possibility has been recently revisited in \citet{Clara} and \citet{Martins} (henceforth Paper 1 and Paper 2, respectively), who show that there is a preference for a value of the fine-structure constant, $\alpha$, at the BBN epoch that is larger than the current laboratory one by a few parts per million of relative variation, even when allowing for possible changes to the most relevant cosmological parameters impacting BBN: the neutron lifetime, the number of neutrino species, the and baryon-to-photon ratio.

Stars are formed with an initial chemical composition representative of their birth environment. During their evolution, this chemical composition is modified by several processes, namely nuclear reactions and transport processes. Nuclear reactions affect the abundance profiles of some elements in the central region of stars, increasing or decreasing their abundances. Transport processes of chemical elements modify the elements distribution in the whole star. There are macroscopic transport processes as, for example, convection, which efficiently transports chemical elements and leads to an homogenized chemical composition in the convective zone. Other macroscopic transport processes are less efficient and reduce the potential internal gradient of chemical composition while transporting them deeper in stars (i.e. transport induced by the rotation: e.g. \citealt[][and references therein]{palacios03,talon08,maeder09}, thermohaline convection: e.g. \citealt[][and references therein]{vauclair04,denissenkov10,brown13,deal16}). These macroscopic transport processes are in competition with microscopic transport processes, namely atomic diffusion \citep[see][for a detailed description]{michaud15}. This comes from physics first principles and is induced by the internal gradients of pressure, temperature and composition. It is efficient in radiative zones and leads to a selective transport of chemical elements, mainly driven by the competition between the gravity and radiative accelerations. Gravity moves elements towards the center of stars while radiative accelerations, a mechanism for transfer of momentum between photons and ions, move some elements toward the surface of stars (depending on their ionisation states and abundances). As a consequence, the surface abundance of elements are either larger or smaller than the initial ones. Considering all the potential processes affecting the chemical elements distribution in stars, it is expected that observed surface abundances are often different from the initial ones, especially on the main sequence where the surface convective zones are not too deep to allow surface abundance variations.

Light elements (such as lithium and beryllium) are especially impacted by transport processes during the evolution of stars. Atomic diffusion leads to a depletion of these elements from the surface (radiative accelerations are most of the time negligible for these elements). Moreover, these elements are destroyed by nuclear reactions at rather low temperatures (about $2.5$ and $3.5$ million~K for lithium and beryllium, respectively), not very deep inside stars. When macroscopic transport processes are efficient enough and affect deep regions, lithium is then transported in regions where the temperature is large enough to be destroyed by nuclear reactions. The combination of both atomic diffusion and macroscopic transport processes is expected to lead to a discernible depletion of lithium from the surface, hence to lithium-7 surface abundances smaller than the initial one \citep[e.g.][]{vauclair88,charbonnel94,Korn,korn07,gruyters13,gruyters16,dumont20}.

Population II stars are not an exception and undergo the same kind of depletion thanks to the same processes \citep[e.g.][]{michaud84,richard05}. This is the reason why the lithium-7 surface abundances observed in population II stars (known as the Spite plateau, \citealt[][]{spite82}) are most likely not the initial ones, hence the discrepancy with a larger cosmological lithium-7 abundance (from BBN). It is interesting to note that a lot of stars are found to have a lithium abundance smaller than the lithium plateau \citep{bonifacio07,cayrel08,sbordone10}, which can also be explained by stellar models, for example, in the case of carbon enhanced metal poor stars with excess in s process elements (CEMP-s stars) \citep[e.g.][]{stancliffe09,deal21}.

Here we draw on the self-consistent perturbative analysis formalism introduced in Paper 1 and extended in Paper 2 to revisit the role of stellar depletion in the Lithium problem, while simultaneously allowing for time variation of nature's fundamental constants, in a broad class of Grand Unified theory (GUT) scenarios where all the gauge and Yukawa couplings are allowed to vary. It will be seen that the impacts of the two mechanisms can be separately constrained. We will start by assuming that the there is a depletion factor relating the cosmological and astrophysical lithium-7 abundances
\be
(^7Li)_{ast}=(1-\Delta)(^7Li)_{cos}\,. \label{depl}
\ee
Throughout our statistical analyses $\Delta$ is allowed to span the entire $[0,1]$ range, under the assumption of a uniform prior. Clearly this is a purely phenomenological parameter, and given any pair or cosmological and astrophysical lithium abundances, for which the former is larger than the latter, there will always be a choice of $\Delta$ that makes the two compatible. In this sense, the observed lithium-7 abundance simply provides a measurement of $\Delta$. However, the hypothesis underlying this assumption is that there exist stellar physics mechanisms and/or physical transport processes occurring in stars that can account for the the relevant value of $\Delta$. Thus our purpose in introducing Eq. (\ref{depl}) is twofold. Firstly, it will allow us to quantify the impact of astrophysical depletion in models which also allow for new physics mechanisms (specifically, in the present work, through varying fundamental constants), thereby comparing the possible roles of the two. Secondly, it provides a convenient way to assess the extent to which stellar physics mechanisms could lead to depletion.

Following the approach of Paper 1 and Paper 2, we have done our statistical likelihood analyses using three different combinations of the four abundances, as follows
\begin{itemize}
    \item The baseline case uses the abundances of helium-4, deuterium and lithium-7, which are the three available cosmological abundances. This will providing the reference values for the best-fit BBN values for $\alpha$.
    \item The extended case adds the helium-3 abundance to the former three; this separation stems from the fact that its observed abundance is a local rather than a cosmological one. In any case, as was pointed out in Paper 1 and Paper 2, this has a negligibly small impact on the derived constraints.
    \item The null case, which uses the helium-4, deuterium and helium-3 abundances but does not use lithium-7. The motivation for this case is that it provides a useful null test of the BBN sensitivity to the value of the fine-structure constant. In other words, if all the relevant physics is the standard one, it is expected that the standard value of $\alpha$ is recovered in this case to some degree of sensitivity that is useful to quantify and compare to other probes, It also provides an indication of the BBN constraints on $\alpha$ on the assumption that the Lithium problem has an astrophysical solution.
\end{itemize}
For convenience, we will use the Baseline, Extended and Null terms to denote each of the three combinations when presenting the results in figures in the early part of the work. (This facilitates comparisons with the results in Paper 1 and  Paper 2.) In the later part, and  when summarizing results in the text and tables throughout the work, we will concentrate on the Baseline case, with Null case results provided for comparison when relevant.

The plan of the rest of this work is as follows. We start in Sect. \ref{cosmo} by phenomenologically quantifying the preferred depletion factor $\Delta$ assuming that the three key cosmological parameters---the neutron lifetime, number of neutrinos and baryon-to-photon ratio---are allowed to vary, but constrained by priors external to BBN. In Sect. \ref{constants}, we report on the analogous study for GUT scenarios where all the gauge and Yukawa couplings are allowed to vary, confirming the previously reported preference for a larger value of $\alpha$ at the BBN epoch. In this case the baryon-to-photon ratio and number of neutrinos are assumed to be fixed at their standard values, while the neutron lifetime is unavoidably affected. Together, these two sections confirm a discrepancy in the values of the baryon-to-photon ratio---or equivalently the primordial Deuterium abundance---preferred by BBN and cosmic microwave background (CMB) plus baryon acoustic oscillation (BAO) data, which has been recently reported \citep{Pitrou20,Yeh}. In Sect. \ref{depletion} we address the Lithium problem, showing that the combination of atomic diffusion, rotation and penetrative convection can bridge the gap between the cosmological abundance and the surface abundances measured in Population II stars. In Sect. \ref{deuterium} we address the Deuterium discrepancy, showing that it (and not the Lithium problem) is driving the aforementioned preference for a larger value of $\alpha$ and further quantifying the model dependence of this preference. Finally, we offer some conclusions in Sect. \ref{concl}.

%%%%%%%%%%%%%%%%%%%%%%%%%%%%%%%%%%%%%%%%%%%%%%%%%%%%%%%%%%%%%%%%%%%%%%%%%%%%%%
\section{Depletion and cosmological parameters}
\label{cosmo}

Generically, the sensitivity of the primordial BBN abundances to the various relevant model parameters can be described as
\begin{equation}
\frac{\Delta Y_i}{Y_i}=\sum_j C_{ij}\frac{\Delta X_j}{X_j}\,,
\end{equation}
where $C_{ij}=\partial\ln{(Y_i)}/\partial\ln{(X_j)}$ are the sensitivity coefficients. The perturbation is always done with respect to the predicted abundance values in some baseline theoretical model, which in our case is the one recently presented in \citet{Pitrou20}. These predicted abundances are listed in Table \ref{table1} for convenience, together with the observed abundances as recommended by the BBN review in \citet{PDG}. The two are compared using standard statistical likelihood methods, and theoretical and observational uncertainties being added in quadrature. We note that our fiducial model differs from the one used in Paper 1 and Paper 2, which was based on the earlier work of \citet{Pitrou}. Therefore our present results can be approximately (but not exactly) compared with those of the earlier papers.

%%%%%%%%%%%%%%%%%%%%%%%%%%%%%%%%%%%%%%%%%%%%%%%%%%%%%%%%%%%%%%%%%%%%%%%%%%%%%%
\begin{table}
\caption{Theoretical and observed primordial abundances used in our analysis. The theoretical ones have been obtained in \citet{Pitrou20}. The observed ones are the recommended values in the 2020 Particle Data Group BBN review, with representative references being, respectively, \citet{Aver}, \citet{Cooke}, \citet{Bania} and \citet{sbordone10}. Note that strictly speaking the ${}^3He$ value is not cosmological; see the main text for further discussion.}
\label{table1}
\centering
\begin{tabular}{c c c}
\hline
Abundance & Theoretical & Observed \\
\hline
$Y_p$ & $0.24721\pm0.00014$ & $0.245\pm0.003$ \\
$(D/H)\times 10^5$ & $2.439\pm0.037$ & $2.547\pm0.025$ \\
$({}^3He/H)\times 10^5$ & $1.039\pm0.014$ & $1.1\pm0.2$ \\
$({}^7Li/H)\times 10^{10}$ & $5.464\pm0.220$ & $1.6\pm0.3$ \\
\hline
\end{tabular}
\end{table}
%%%%%%%%%%%%%%%%%%%%%%%%%%%%%%%%%%%%%%%%%%%%%%%%%%%%%%%%%%%%%%%%%%%%%%%%%%%%%%

In this section we consider the effect of the values of the neutron lifetime $\tau_n$, the number of neutrino species $N_\nu$, and the baryon-to-photon ratio $\eta_{10}$ (where, for convenience, we have defined $\eta_{10}=\eta\times 10^{10}$) on the BBN abundances. The fiducial values of these parameters (which will be used as statistical Gaussian priors in the analysis) are, respectively
\be
\tau_n=879.4\pm0.6\, s
\ee
for the neutron lifetime \citep{PDG} (see also \citet{Rajan} for a recent discussion of experimental measurements of this lifetime),
\be
N_\nu=2.984\pm0.008
\ee
for the number of neutrino species, which comes from the LEP measurement \citep{PDG}, and
\be
\eta_{10}=6.143\pm0.038\,,
\ee
from the combination of recent CMB (Planck 2018) and BAO data \citep{Planck}. For reference, the latter corresponds to the physical baryon density
\be
\omega_b=\Omega_bh^2=\frac{\eta_{10}}{274}=0.02242\pm0.00014\,.
\ee

%%%%%%%%%%%%%%%%%%%%%%%%%%%%%%%%%%%%%%%%%%%%%%%%%%%%%%%%%%%%%%%%%%%%%%%%%%%%%%
\begin{table}
\caption{Sensitivity coefficients of BBN nuclide abundances on the cosmological parameters in our phenomenological parametrisation, defined in the main text.}
\label{table2}
\centering
\begin{tabular}{| c | c c c c |}
\hline
$C_{ij}$ & D & ${}^3$He & ${}^4$He & ${}^7$Li \\
\hline
$t_i$ & +0.442 & +0.141 & +0.732 & +0.438 \\
$v_i$ & +0.409 & +0.136 & +0.164 & -0.277 \\
$w_i$ & -1.65 & -0.567 & +0.039 & +2.08 \\
\hline
\end{tabular}
\end{table}
%%%%%%%%%%%%%%%%%%%%%%%%%%%%%%%%%%%%%%%%%%%%%%%%%%%%%%%%%%%%%%%%%%%%%%%%%%%%%%

With these assumptions our perturbative analysis for the BBN abundances has the specific form
\begin{equation}
\frac{\Delta Y_i}{Y_i}=t_i\frac{\Delta\tau_n}{\tau_n}+v_i\frac{\Delta N_\nu}{N_\nu}+w_i\frac{\Delta\eta}{\eta}\,,
\end{equation}
where $t_i$, $v_i$ and $w_i$ are the sensitivity coefficients listed in Table \ref{table2} and previously discussed in Paper 2 and references therein. Note that Deuterium is far more sensitive to the baryon fraction than to the neutron lifetime or the number of neutrinos, while the opposite occurs for Helium-4. Additionally, we will make the assumption that the cosmological and astrophysical lithium-7 abundances are related through Eq. (\ref{depl}), so the depletion factor $\Delta$ becomes a fourth model parameter, together with the relative variations of the three cosmological parameters.

%%%%%%%%%%%%%%%%%%%%
\begin{figure*}
\centering
\includegraphics[width=0.3\textwidth]{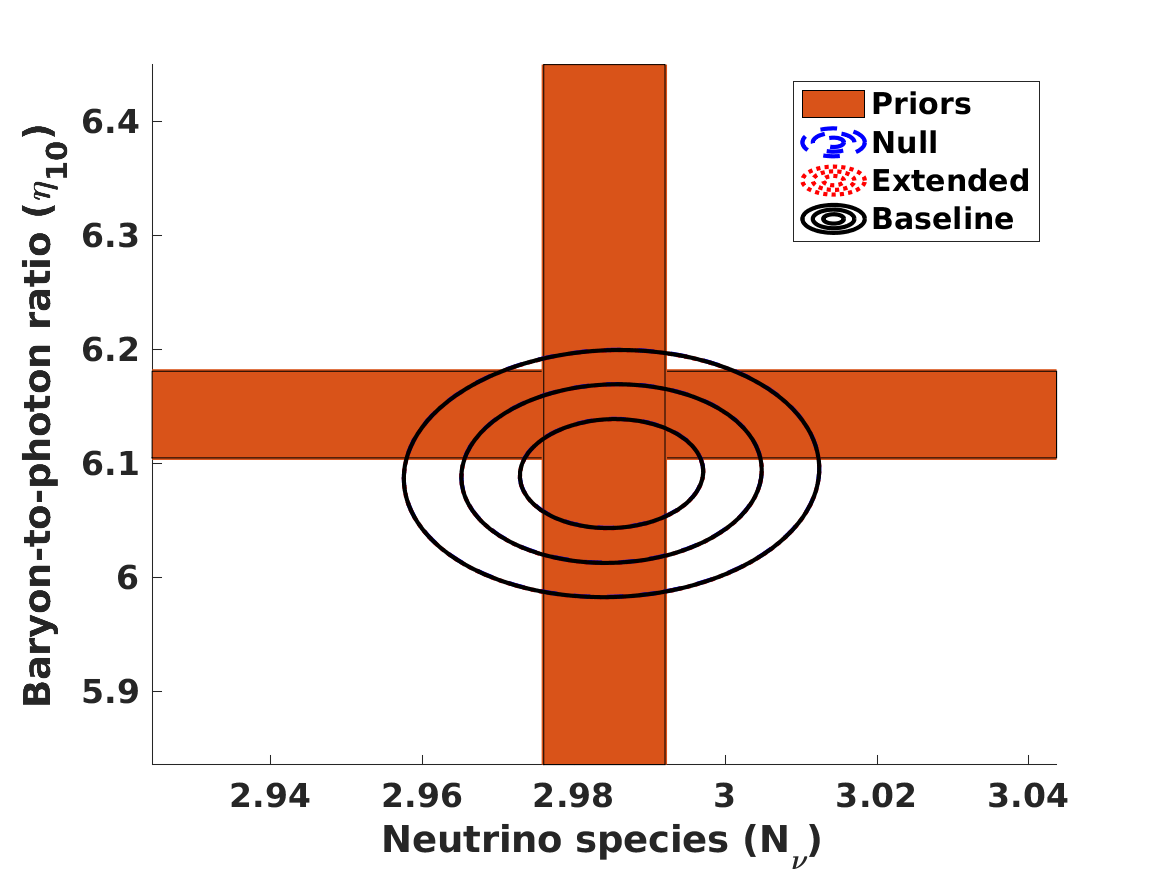}
\includegraphics[width=0.3\textwidth]{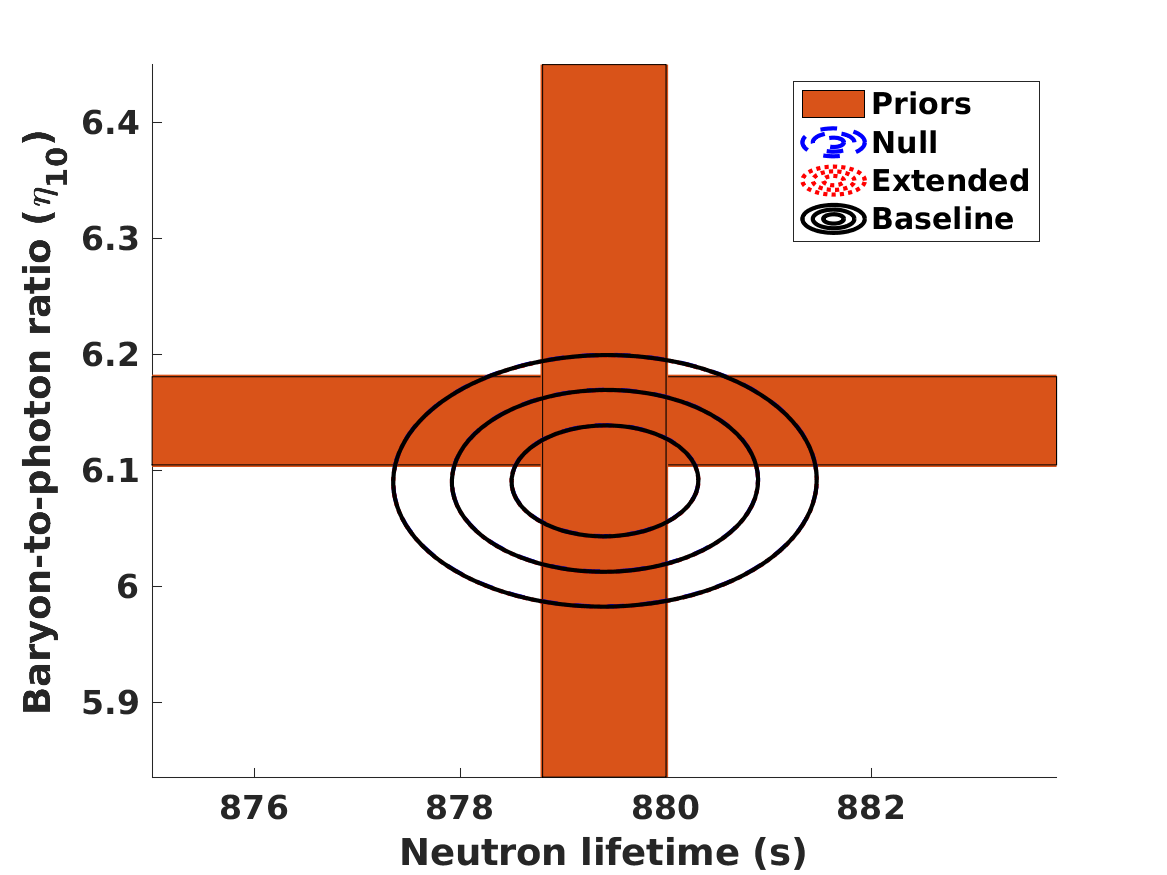}
\includegraphics[width=0.3\textwidth]{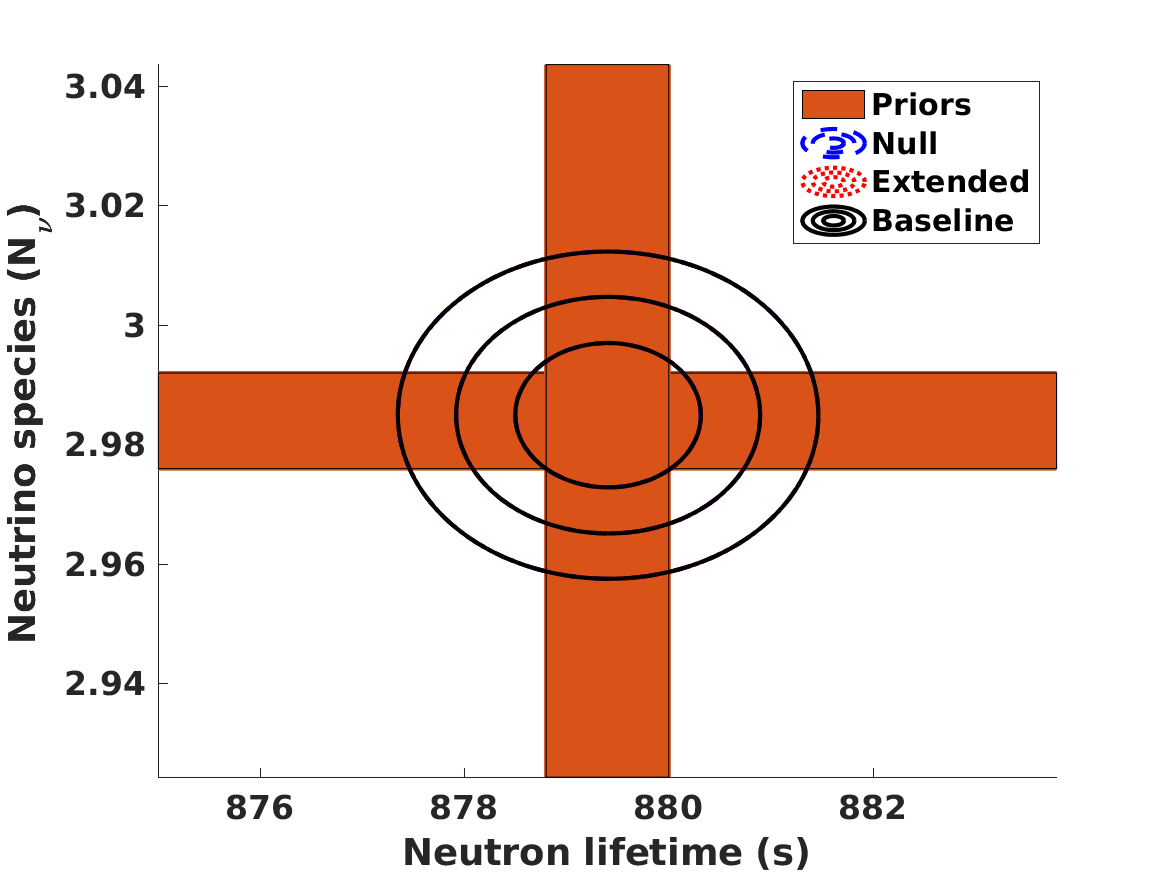}
\includegraphics[width=0.3\textwidth]{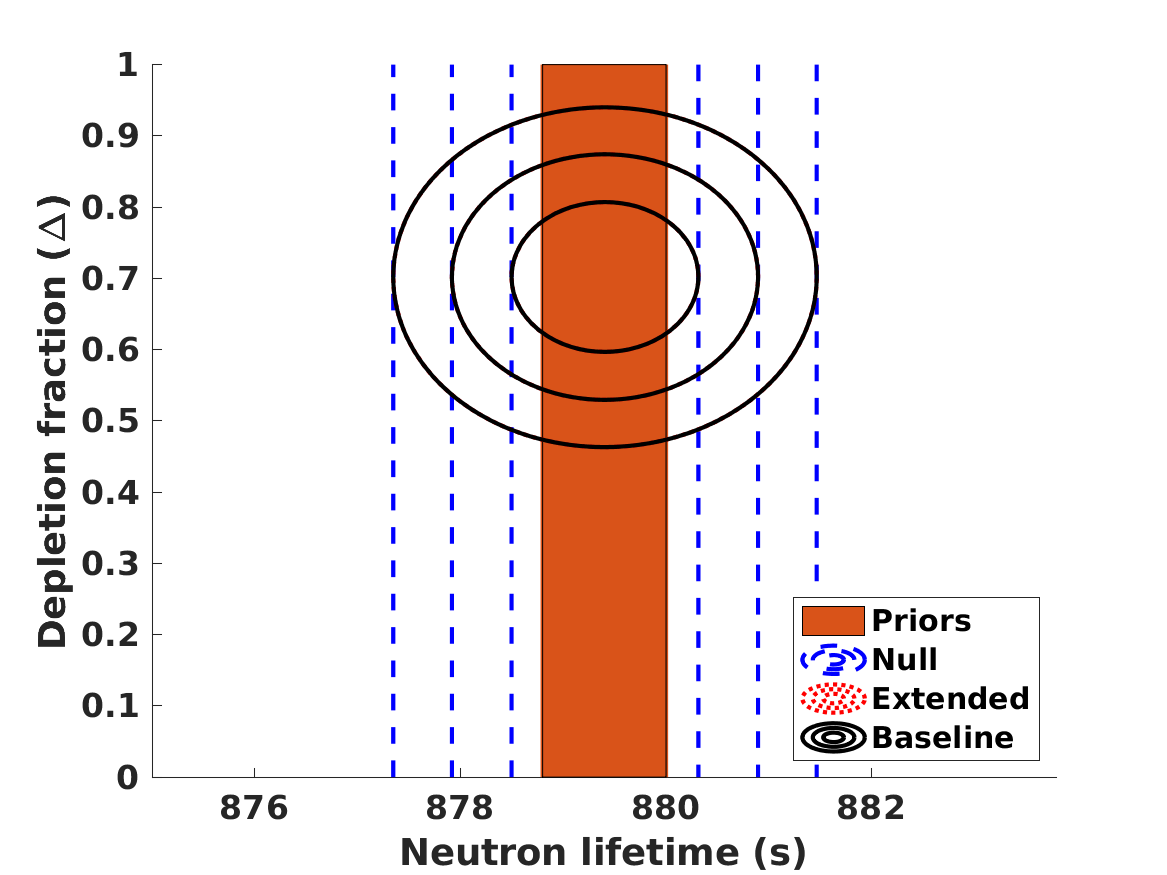}
\includegraphics[width=0.3\textwidth]{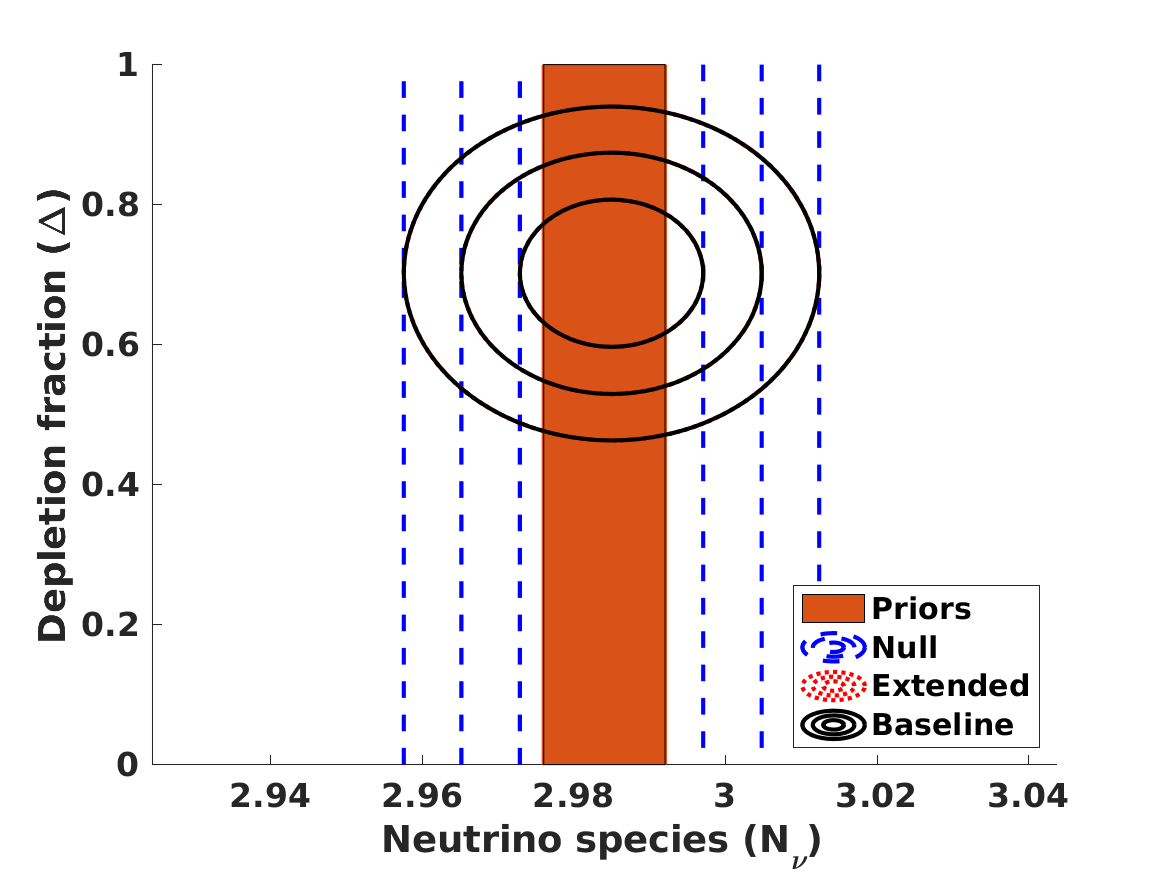}
\includegraphics[width=0.3\textwidth]{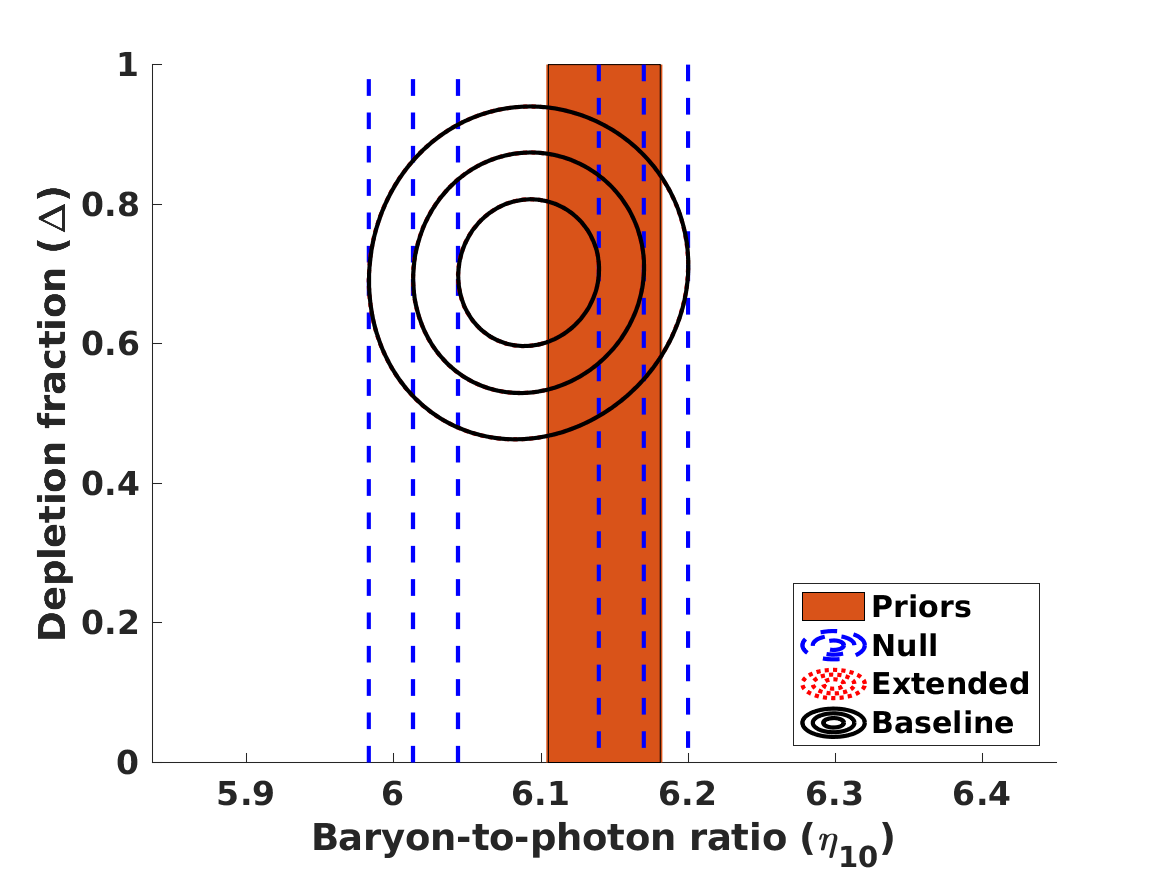}
\includegraphics[width=0.23\textwidth]{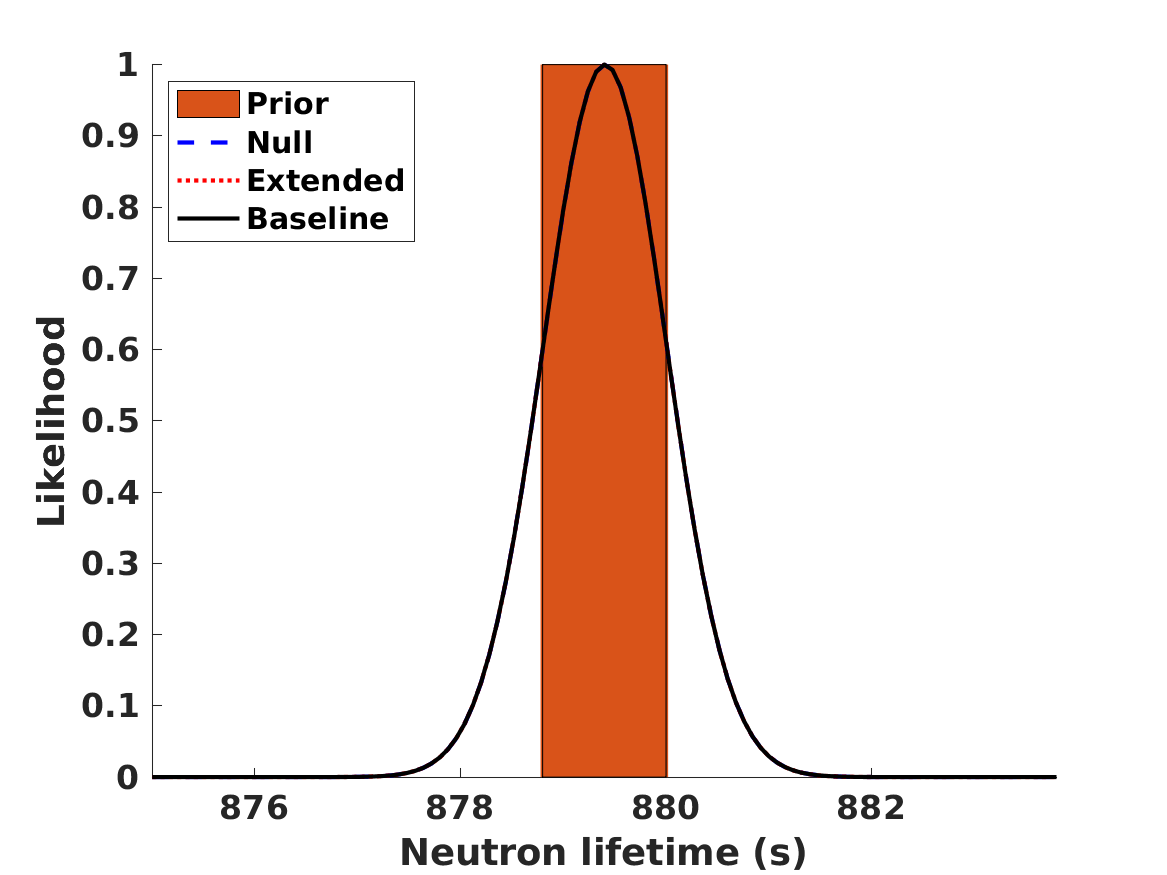}
\includegraphics[width=0.23\textwidth]{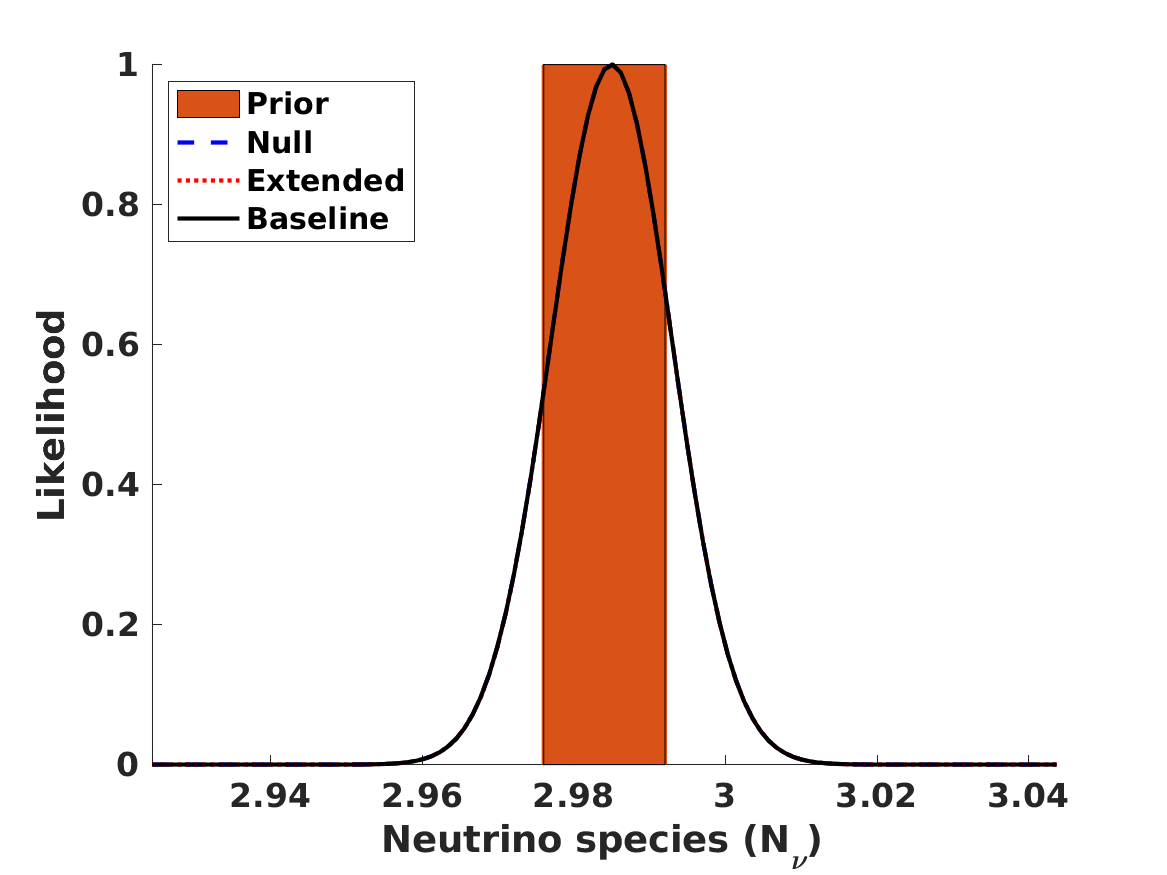}
\includegraphics[width=0.23\textwidth]{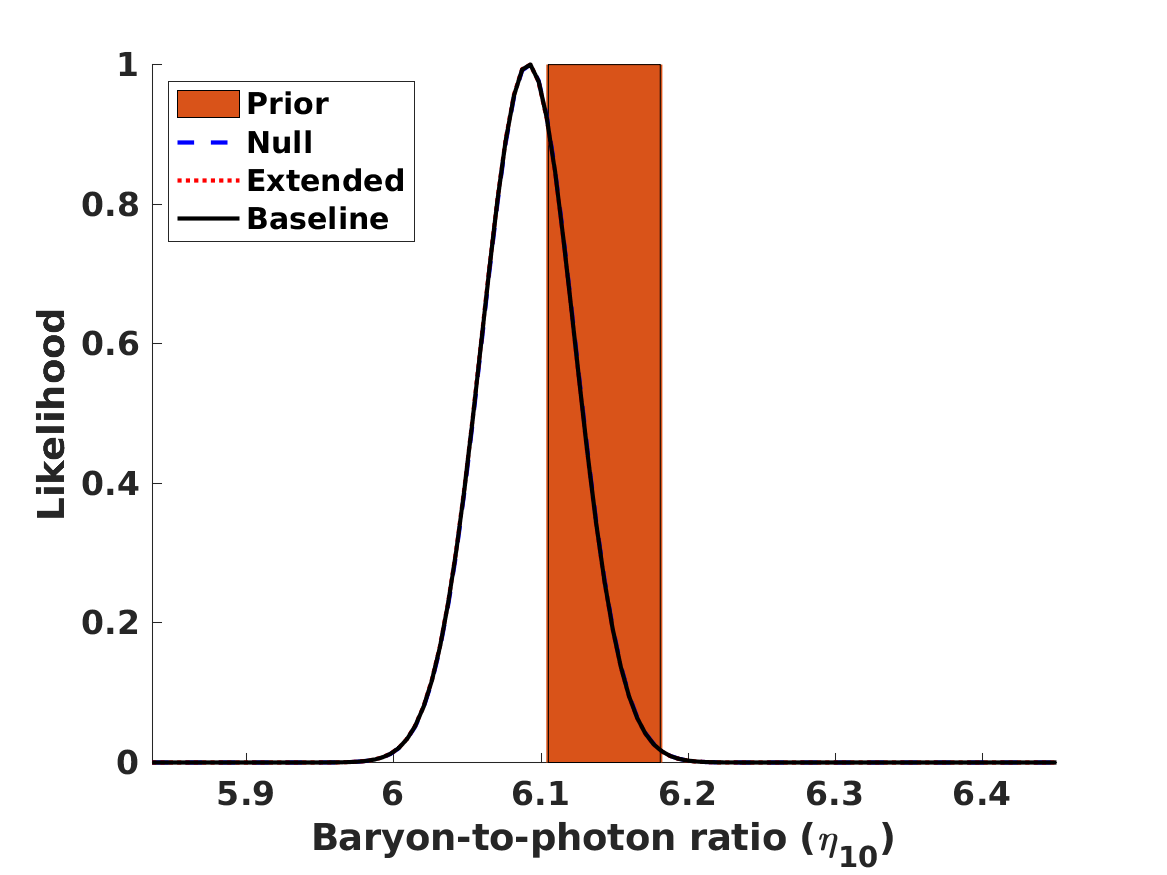}
\includegraphics[width=0.23\textwidth]{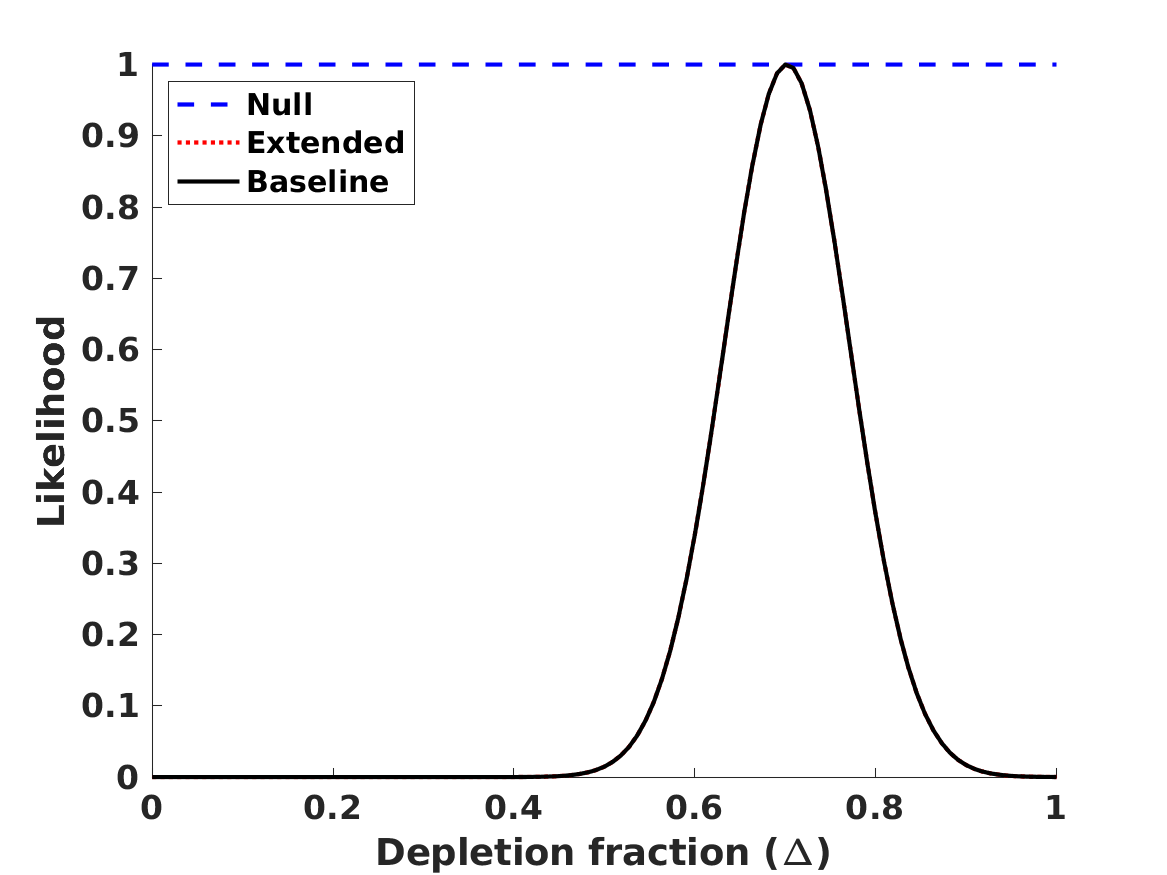}
\caption{BBN constraints on the lithium-7 phenomenological depletion factor and key cosmological parameters, with remaining parameters marginalized in each case. The one-sigma range corresponding to the priors on cosmological parameters mentioned in the text are also shown for illustration purposes. The two-dimensional planes depict the 68.3, 95.4 and 99.7 percent confidence levels.}
\label{figure1}
\end{figure*}
%%%%%%%%%%%%%%%%%%%%

The results of the analysis are depicted in Figure \ref{figure1}. The best-fit values of our free parameters (with their one standard deviation uncertainties) in the Baseline case are
\bq
\frac{\Delta\tau_n}{\tau_n}&=&0.0000\pm0.0007\\
\frac{\Delta N_\nu}{N_\nu}&=&0.0004\pm0.0022\\
\frac{\Delta\eta_{10}}{\eta_{10}}&=&-0.0082\pm0.0050\\
\Delta&=&0.70\pm0.07\,;
\eq
in the Null case the first three of these are the same, while $\Delta$ is obviously not constrained. (On the other hand, if one assumes that $\tau_n$, $N_\nu$ and $\eta_{10}$ are fixed at their fiducial values rather than allowed to vary and marginalized, one finds $\Delta=0.71\pm0.07$.) These lead to the following best fit theoretical abundances
\bq
Y_p&=&0.247\pm0.001\\
(D/H)\times 10^5&=&2.474\pm0.043\\
({}^3He/H)\times 10^5&=&1.044\pm0.015\\
({}^7Li/H)\times 10^{10} (Cos)&=&5.370\pm0.224\\
({}^7Li/H)\times 10^{10} (Ast)&=&1.611\pm0.387\,;\\
\eq
for lithium-7 we separately list the cosmological (primordial) abundance, and the corresponding astrophysical one, corrected by the preferred seventy percent depletion. We confirm the result, already mentioned by \citet{Pitrou20}, that with the latest nuclear physics cross sections and observed abundances the BBN observations prefer a slightly lower $\eta_{10}$ than the one measured by the combination of CMB and BAO observations. Alternatively, this can be phrased as as discrepancy in the Deuterium abundance. On the other hand, different reaction rates are used in \citet{Yeh} and in \citet{Pisanti}, which find no such discrepancy. In any case, as illustrated in Figure \ref{figure1}, the discrepancy is only slightly larger than one standard deviation, so its statistical significance is limited. We revisit this point later in this work.

%%%%%%%%%%%%%%%%%%%%%%%%%%%%%%%%%%%%%%%%%%%%%%%%%%%%%%%%%%%%%%%
\section{Depletion and Varying fundamental constants}
\label{constants}

The approach introduced Paper 1 and Paper 2 to study BBN with varying fundamental constants is also a perturbative one. It builds upon earlier work by \citet{Muller} and \citet{Resonance1} and subsequent developments by \citet{Coc} and \citet{Stern1}, to generically write the relative variations of other couplings as the product of some constant particle physics coefficients and the relative variation of the fine-structure constant. With some reasonable simplifying assumptions, which stem from the work of \citet{Campbell}, only two such dimensionless coefficients are needed, one pertaining to electroweak physics (denoted $S$) and the other to strong interactions (denoted $R$). This enables a phenomenological description of a broad range of GUT  scenarios. We refer the reader to Paper 1 and Paper 2 for a more detailed description of this formalism. In what follows, astrophysical constraints on $\alpha$ are expressed in terms of a relative variation with respect to the local laboratory value, 
\be
\frac{\Delta\alpha}{\alpha}(z)\equiv\frac{\alpha(z)-\alpha_0}{\alpha_0}\,,
\ee
where $\alpha_0=0.0072973525693(11)$ is the local laboratory value.

Under these assumptions, the sensitivity of the primordial BBN abundances to the values of the fundamental constants will therefore be expressed as a function of the phenomenological particle physics parameters $R$ and $S$ and the relative variation of $\alpha$, in other words
\begin{equation}\label{paramalpha}
\frac{\Delta Y_i}{Y_i}=(x_i+y_iS+z_iR)\frac{\Delta\alpha}{\alpha}\,;
\end{equation}
the corresponding sensitivity coefficients are listed in Table \ref{table3}. Note that the baryon-to-photon ratio and number of neutrinos are assumed to be fixed at their standard values. On the other hand the neutron lifetime will be affected by the $\alpha$ variation, according to
\begin{equation}
\frac{\Delta \tau_n}{\tau_n}=[-0.2-2.0~S+3.8~R]\, \frac{\Delta\alpha}{\alpha}\,,
\end{equation}
but this effect has been included in the computation of these sensitivity coefficients, so the neutron lifetime does not appear as an explicit parameter. We refer the reader to Paper 2 for a more detailed discussion.

%%%%%%%%%%%%%%%%%%%%%%%%%%%%%%%%%%%%%%%%%%%%%%%%%%%%%%%%%%%%%%%%%%%%%%%%%%%%%%
\begin{table}
\caption{Sensitivity coefficients of BBN nuclide abundances on the free unification parameters of our phenomenological parametrisation of GUT scenarios, defined in the main text. The baryon-to-photon ratio and number of neutrinos are assumed to be fixed at their standard values.}
\label{table3}
\centering
\begin{tabular}{| c | c c c c |}
\hline
$C_{ij}$ & D & ${}^3$He & ${}^4$He & ${}^7$Li \\
\hline
$x_i$ & +42.0 & +1.27 & -4.6 & -166.6 \\
$y_i$ & +39.2 & +0.72 & -5.0 & -151.6 \\
$z_i$ & +36.6 & -89.5 & +14.6 & -200.9 \\
\hline
\end{tabular}
\end{table}
%%%%%%%%%%%%%%%%%%%%%%%%%%%%%%%%%%%%%%%%%%%%%%%%%%%%%%%%%%%%%%%%%%%%%%%%%%%%%%

The phenomenological parameters $R$ and $S$ can in principle be taken as free parameters, to be experimentally or observationally constrained. Our current knowledge of particle physics and unification scenarios suggests that their absolute values can be anything from order unity to several hundreds, with $R$ allowed to be positive or negative (though with the former case being the more likely one), while $S$ is expected to be non-negative.

The analysis of Paper 1 and Paper 2 has identified three models that can provide a solution to the Lithium problem (at least in the sense of making all theoretical and observed abundances agree to within three standard deviations or less) for the best-fit values of $\Delta\alpha/\alpha$ listed in Table \ref{table4}. Note that the fiducial model used in these papers is not the same as the one used in the present work, and that the full parameter space was different in the two papers.

%%%%%%%%%%%%%%%%%%%%%%%%%%%%%%%%%%%%%%%%%%%%%%%%%%%%%%%%%%%%%%%%%%%%%%%%%%%%%%
\begin{table}
\caption{Constraints on $\Delta\alpha/\alpha$, in the Baseline and Null scenarios, for the Unification, Dilaton and Clocks models, as reported in Paper 1 \citep{Clara} and Paper 2 \citep{Martins}. The listed values are in ppm, and correspond to the best fit in each case and to the range of values within $\Delta\chi^2=4$ of it. Note that the fiducial model used in these papers is not the same as the one used in the present work, and that the full parameter space was different in the two papers.}
\label{table4}
\centering
\begin{tabular}{c c c c}
\hline
Scenarios & Unification & Dilaton & Clocks \\
\hline
Paper 1 (Baseline)  & $12.5\pm2.9$ & $19.9\pm4.5$ & $2.2^{+15.6}_{-0.6}$ \\
Paper 1 (Null)  & $4.6\pm3.8$ & $5.8\pm6.5$ & $1.0^{+7.1}_{-0.9}$ \\
\hline
Paper 2 (Baseline) & $16.1\pm3.9$ & $22.6\pm5.5$ & $2.5^{+3.5}_{-0.5}$ \\
Paper 2 (Null) & $4.6\pm5.5$ & $3.7\pm8.4$ & $0.8^{+1.5}_{-0.6}$ \\
\hline
\end{tabular}
\end{table}
%%%%%%%%%%%%%%%%%%%%%%%%%%%%%%%%%%%%%%%%%%%%%%%%%%%%%%%%%%%%%%%%%%%%%%%%%%%%%%

%%%%%%%%%%%%%%%%%%%%
\begin{figure*}
\centering
\includegraphics[width=0.3\textwidth]{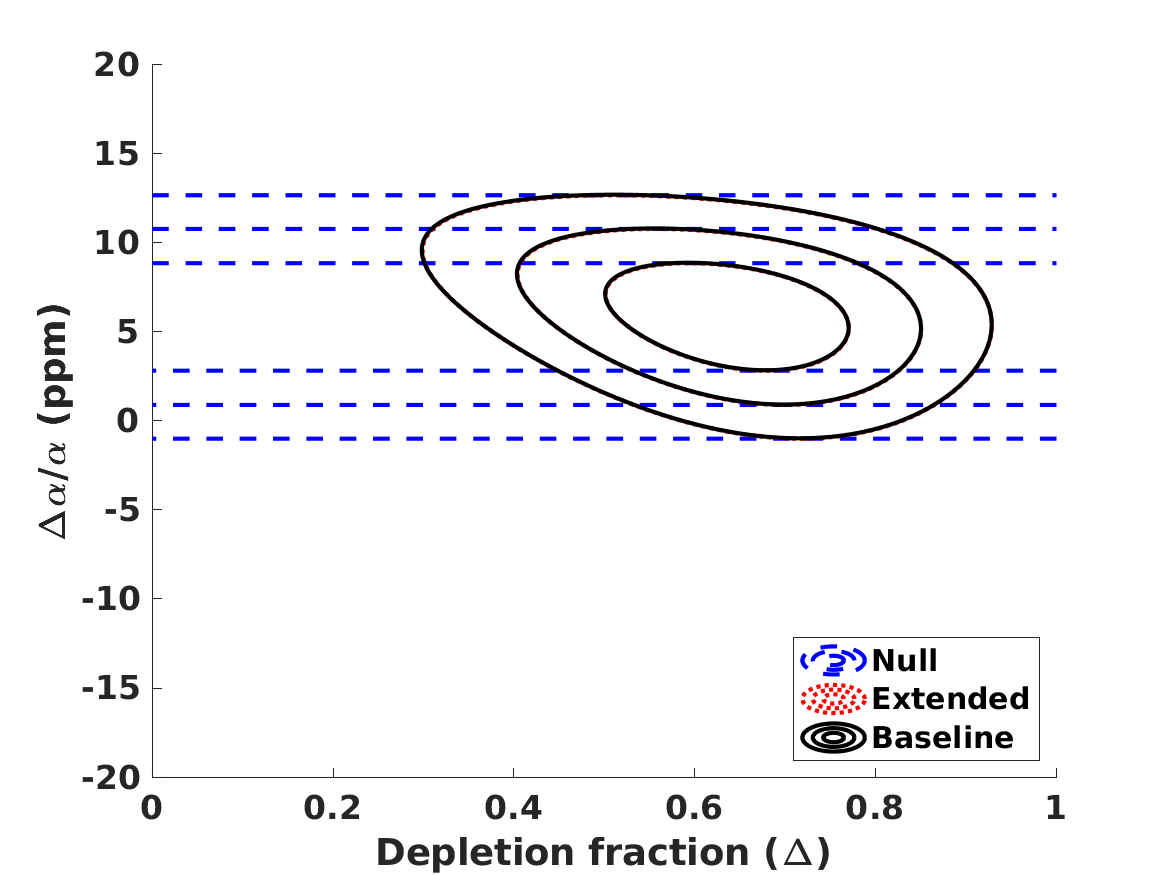}
\includegraphics[width=0.3\textwidth]{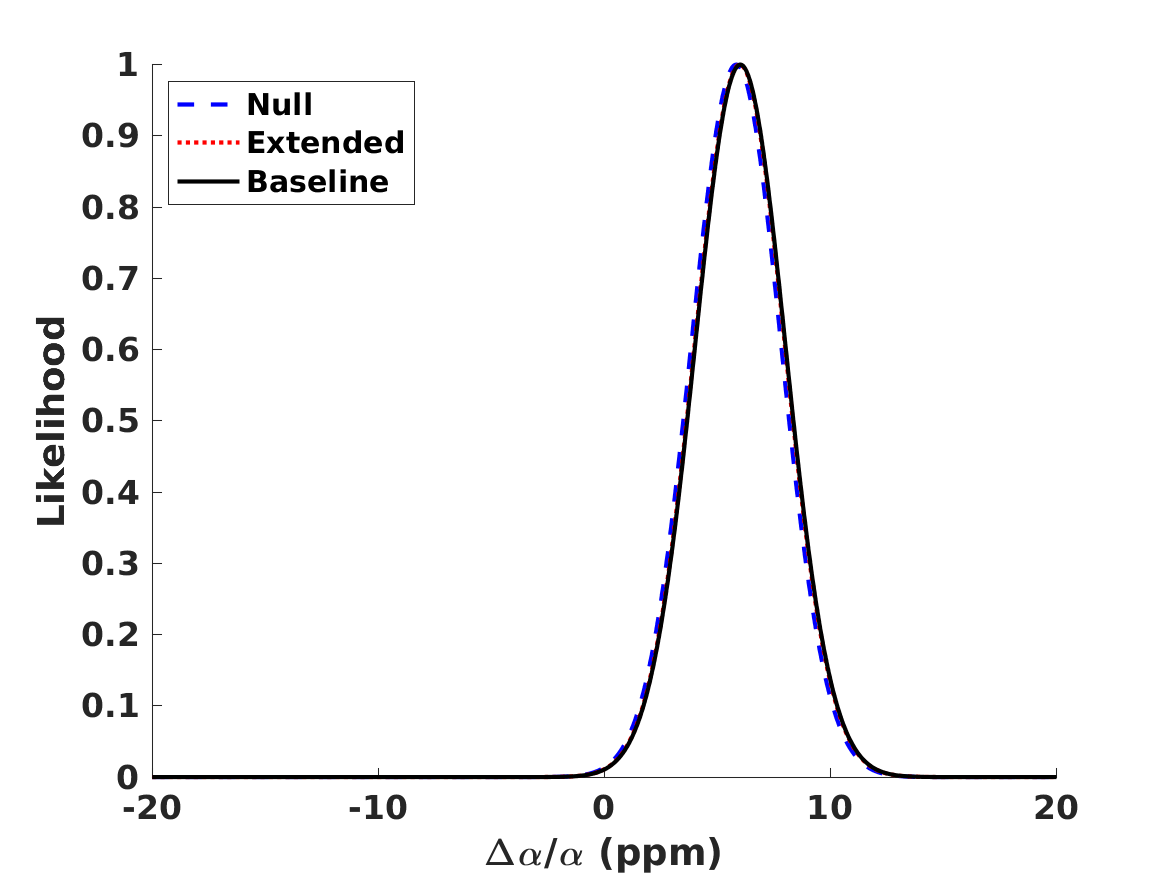}
\includegraphics[width=0.3\textwidth]{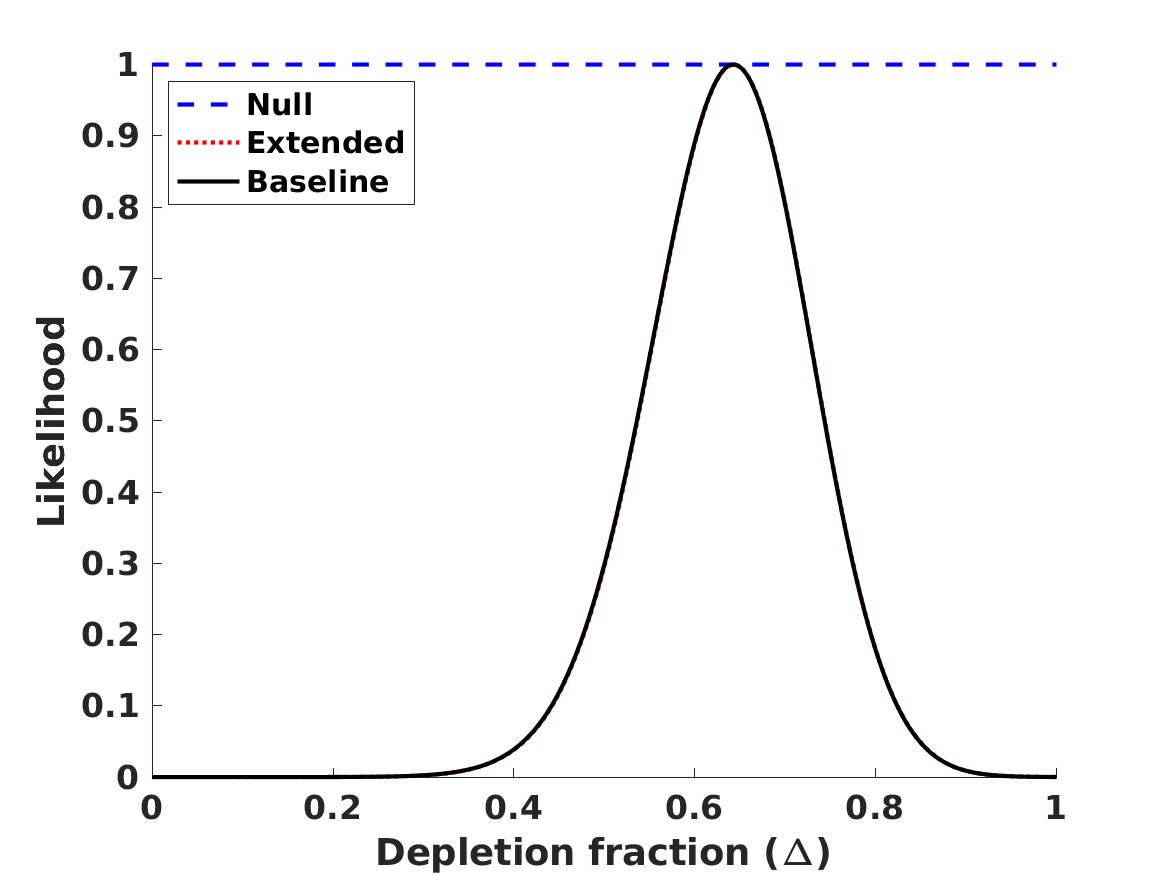}
\includegraphics[width=0.3\textwidth]{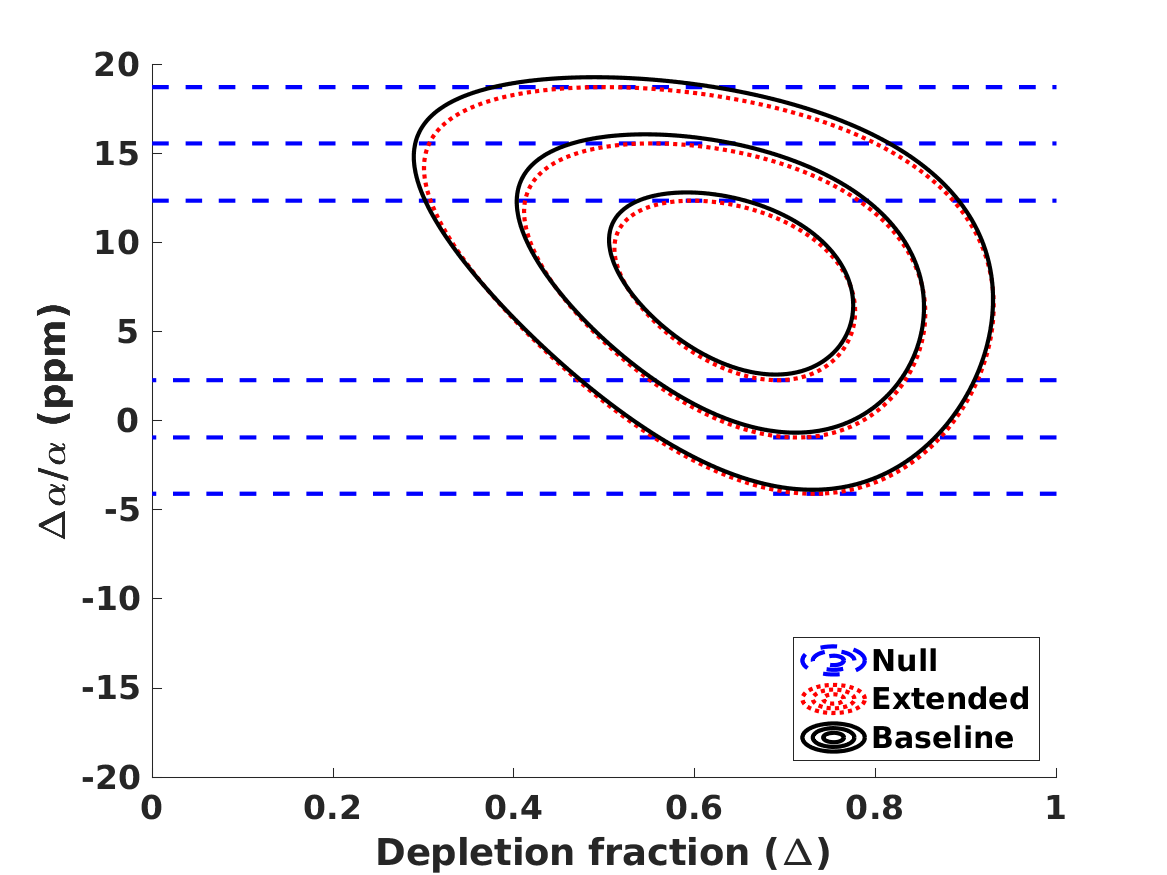}
\includegraphics[width=0.3\textwidth]{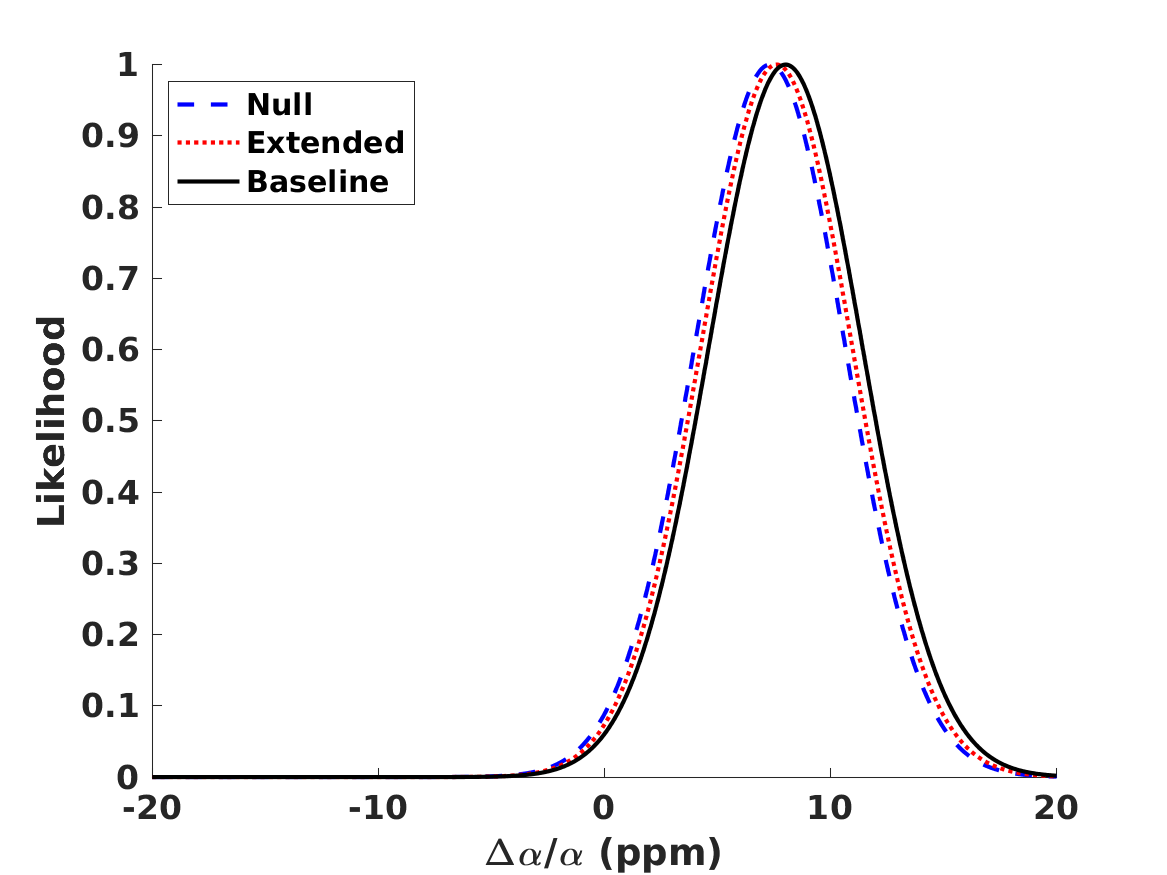}
\includegraphics[width=0.3\textwidth]{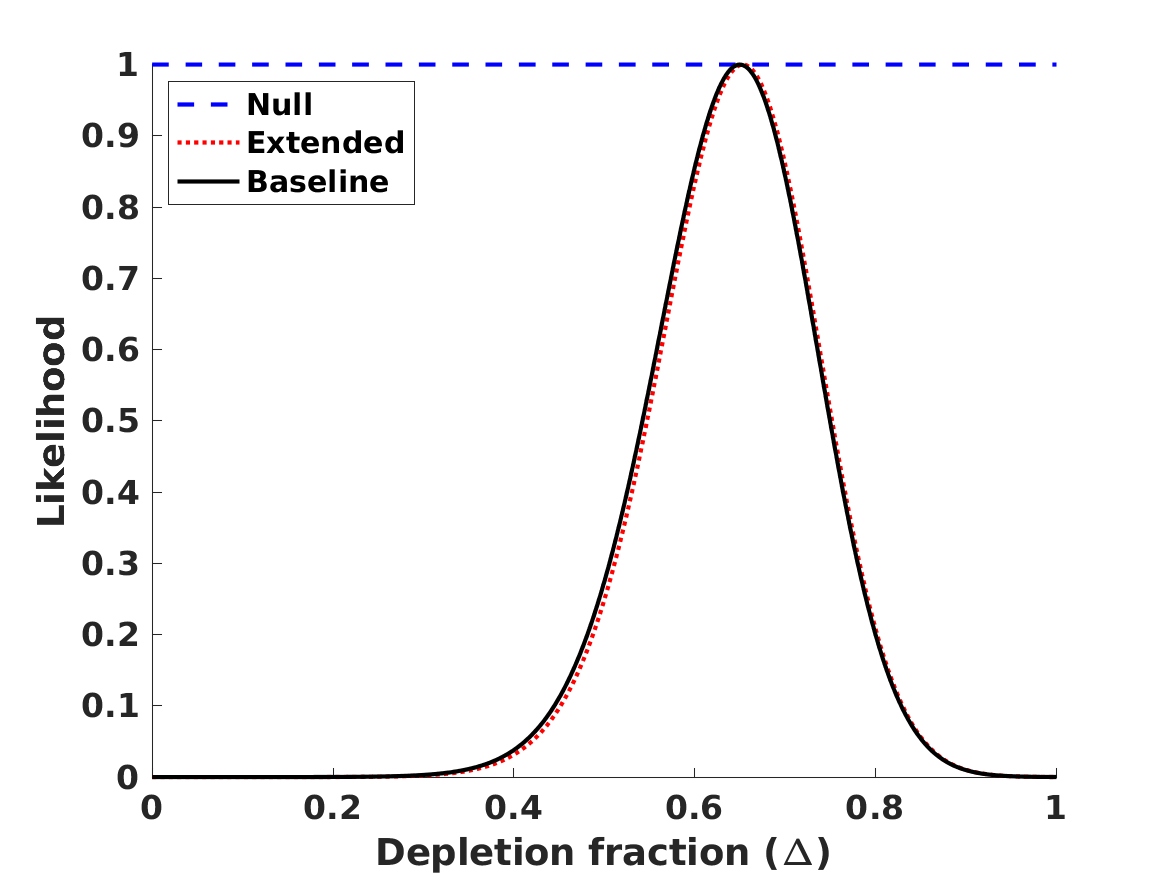}
\includegraphics[width=0.3\textwidth]{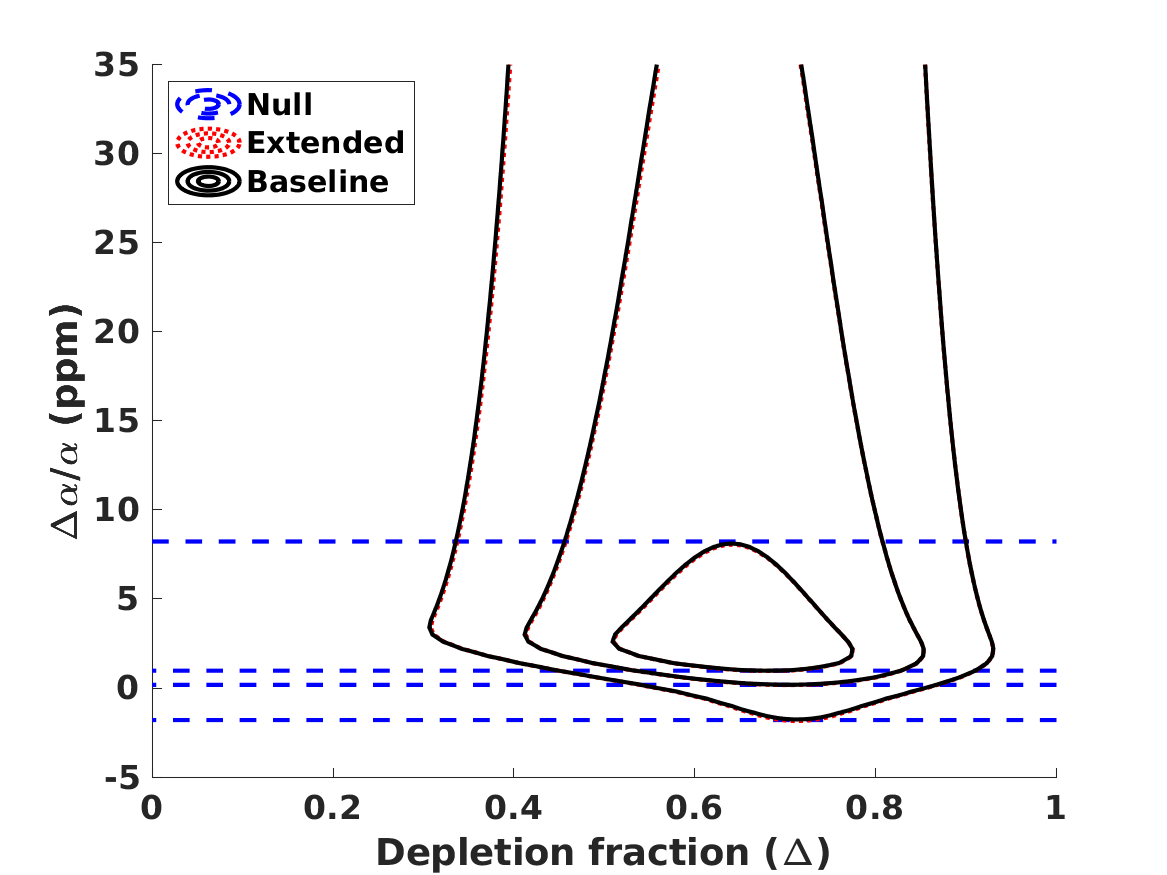}
\includegraphics[width=0.3\textwidth]{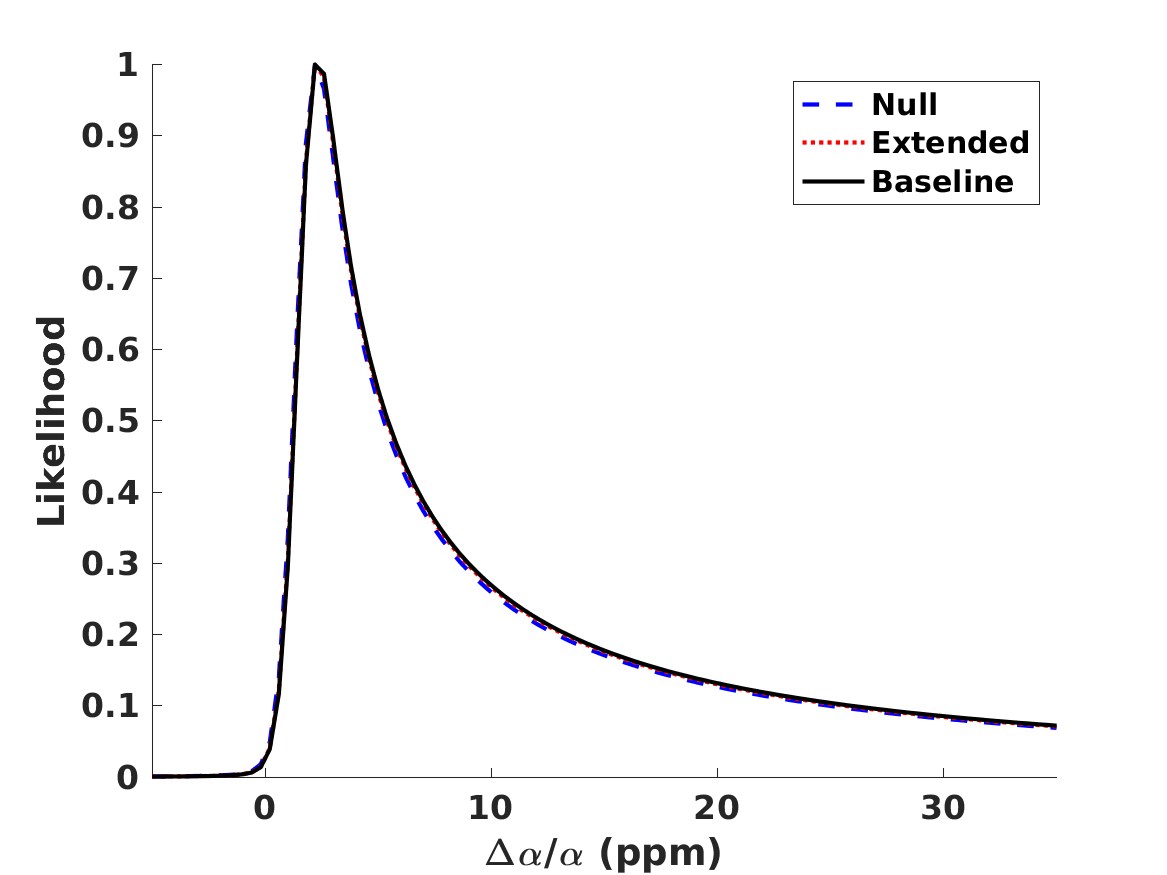}
\includegraphics[width=0.3\textwidth]{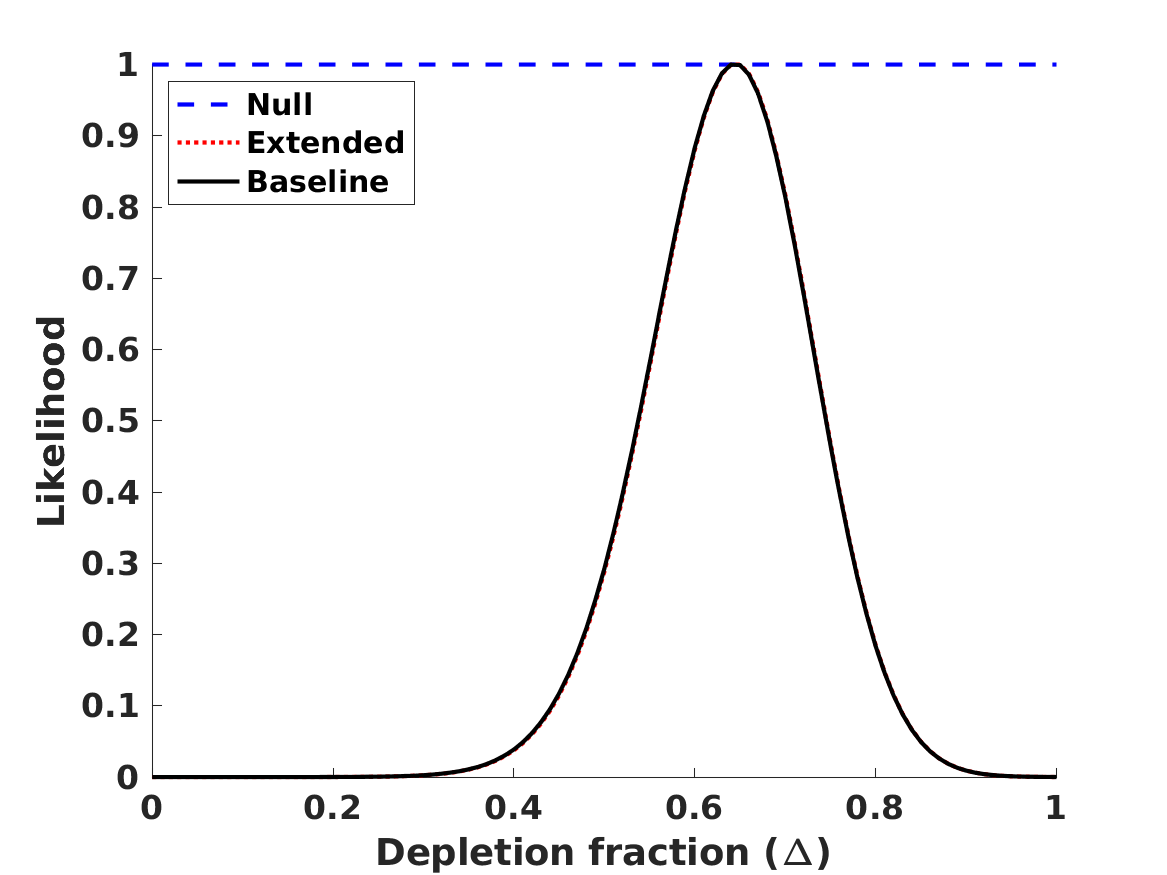}
\caption{BBN constraints on the fine-structure constant and the lithium-7 phenomenological depletion factor, with remaining parameters marginalized in each case. The two-dimensional planes depict the 68.3, 95.4 and 99.7 percent confidence levels. Panels in the first, second and third rows correspond to the Unification, Dilaton and Clocks models.}
\label{figure2}
\end{figure*}
%%%%%%%%%%%%%%%%%%%%

The first of these is a `typical' unification scenario, for which one has \citep{Coc,Langacker}
\begin{equation}
R\sim36\,,\quad S\sim160\,;
\end{equation}
we refer to this as the Unification model. The second is the dilaton-type model discussed by \citet{Nakashima}, for which
\begin{equation}
R\sim109.4\,,\quad S\sim0\,;
\end{equation}
we refer to this as the Dilaton model. Finally the Clocks model denotes the general case where the parameters $R$ and $S$ are allowed to vary and are then marginalised, in this case with uniform priors in the range $R=[0,+500]$, $S=[0,+1000]$, together with an additional prior coming from local experiments with atomic clocks \citep{Clocks}
\begin{equation}\label{clopri}
(1+S)-2.7R=-5\pm15\,.
\end{equation}
This is therefore a more phenomenological model than the previous two (and has a larger number of free parameters than them), but serves the purpose of illustrating the range of behaviours that might be found in GUT models---although we should mention that, unsurprisingly, the results in this case will also depend on the choice of priors for $R$ and $S$. This last point has been discussed in Paper 1.

We now repeat the statistical likelihood analysis, allowing for both a variation of fundamental couplings and lithium-7 depletion, for these three models. In this section we assume that the values of the number of neutrino species and the baryon-to-photon ratio are fixed at their standard fiducial values, but the neutron lifetime is implicitly allowed to vary, as pointed out above. The results of this analysis are depicted in Figure \ref{figure2} and also summarized in Table \ref{table5} for the Baseline case; naturally, in the Null case the depletion fraction is unconstrained, but for comparison we also list the value of $\Delta\alpha/\alpha$ in that case.

%%%%%%%%%%%%%%%%%%%%%%%%%%%%%%%%%%%%%%%%%%%%%%%%%%%%%%%%%%%%%%%%%%%%%%%%%%%%%%
\begin{table*}
\caption{Constraints on $\Delta\alpha/\alpha$ and $\Delta$ for the Unification, Dilaton and Clocks models (with the range of values within $\Delta\chi^2=1$ of it, corresponding to the $68.3\%$ confidence level for a Gaussian posterior likelihood), together with the derived nuclide abundances for each of the best-fit models. For lithium-7 we separately list the cosmological (primordial) abundance, and the corresponding astrophysical one, corrected by the preferred depletion. The results are for the Baseline scenario, with the exception of the first row, which shows the value of $\Delta\alpha/\alpha$ in the Null case for comparison.}
\label{table5}
\centering
\begin{tabular}{| c | c | c | c |}
\hline
Parameter  & Unification & Dilaton & Clocks \\
\hline
$\frac{\Delta\alpha}{\alpha}$ (ppm), Null & $5.8\pm2.0$ & $7.3\pm3.3$ & $2.2^{+2.4}_{0.8}$ \\
\hline
$\frac{\Delta\alpha}{\alpha}$ (ppm), Baseline & $6.0\pm2.0$ & $8.0\pm3.4$ & $2.2^{+2.4}_{0.8}$ \\
$\Delta$ & $0.64\pm0.09$ & $0.65\pm0.09$ & $0.64\pm0.09$\\
\hline
$Y_p$ & $0.247\pm0.001$ & $0.249\pm0.001$ & $0.247\pm0.002$ \\
$(D/H)\times 10^5$ & $2.55\pm0.05$ & $2.52\pm0.05$ & $2.45\pm0.07$ \\
$({}^3He/H)\times 10^5$ & $1.02\pm0.02$ & $0.96\pm0.04$ & $1.04\pm0.07$ \\
$({}^7Li/H)\times 10^{10}$ (Cos) & $4.42\pm0.25$ & $4.51\pm0.24$ & $5.36\pm0.26$ \\
$({}^7Li/H)\times 10^{10}$ (Ast) & $1.59\pm0.52$ & $1.58\pm0.52$ & $1.92\pm0.60$ \\
\hline
\end{tabular}
\end{table*}
%%%%%%%%%%%%%%%%%%%%%%%%%%%%%%%%%%%%%%%%%%%%%%%%%%%%%%%%%%%%%%%%%%%%%%%%%%%%%%

%%%%%%%%%%%%%%%%%%%%
\begin{figure*}
\centering
\includegraphics[width=0.45\textwidth]{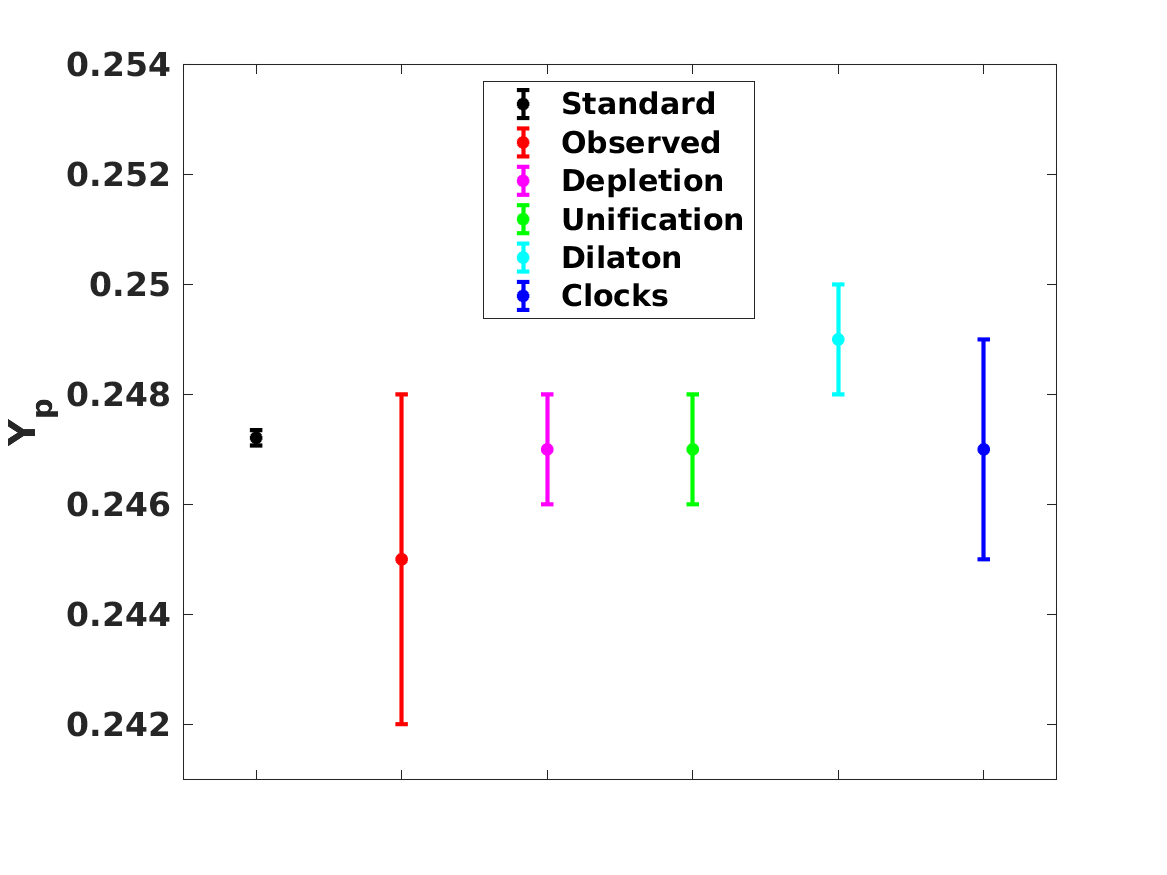}
\includegraphics[width=0.45\textwidth]{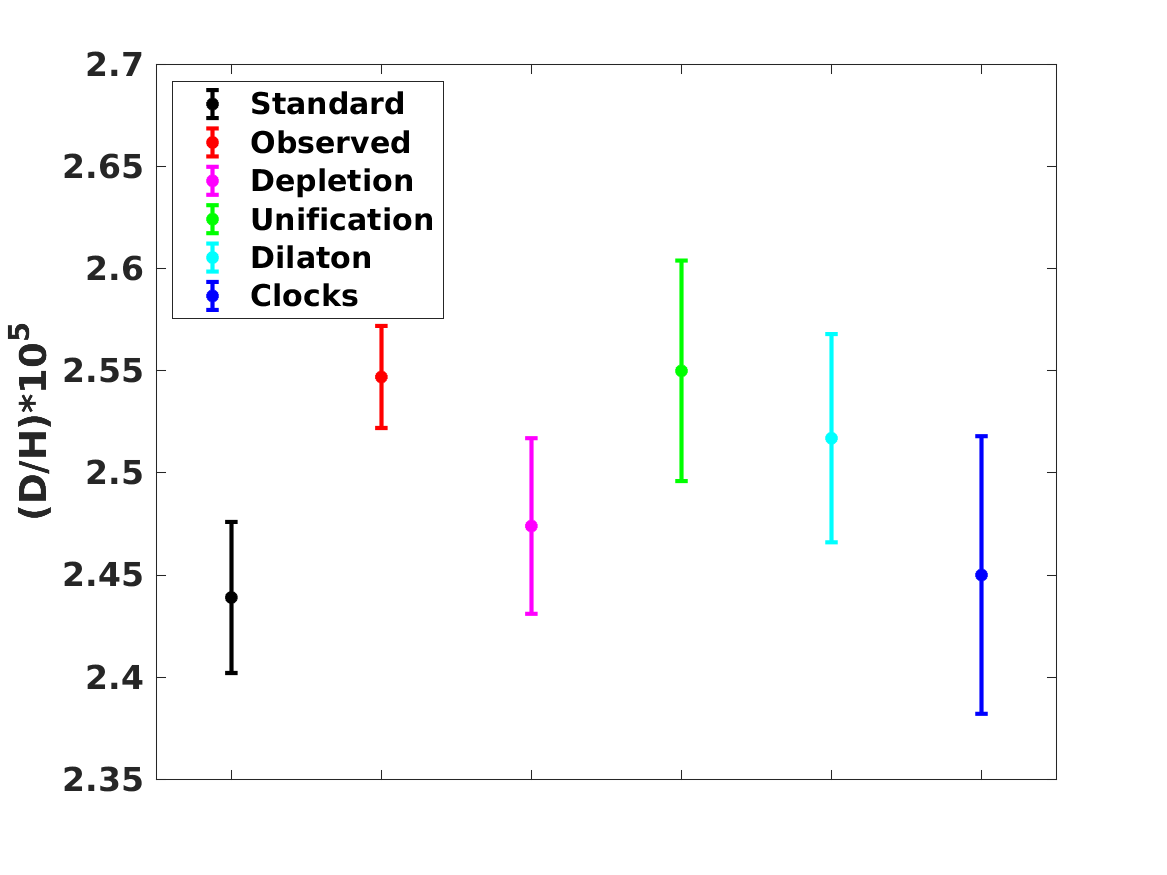}
\includegraphics[width=0.45\textwidth]{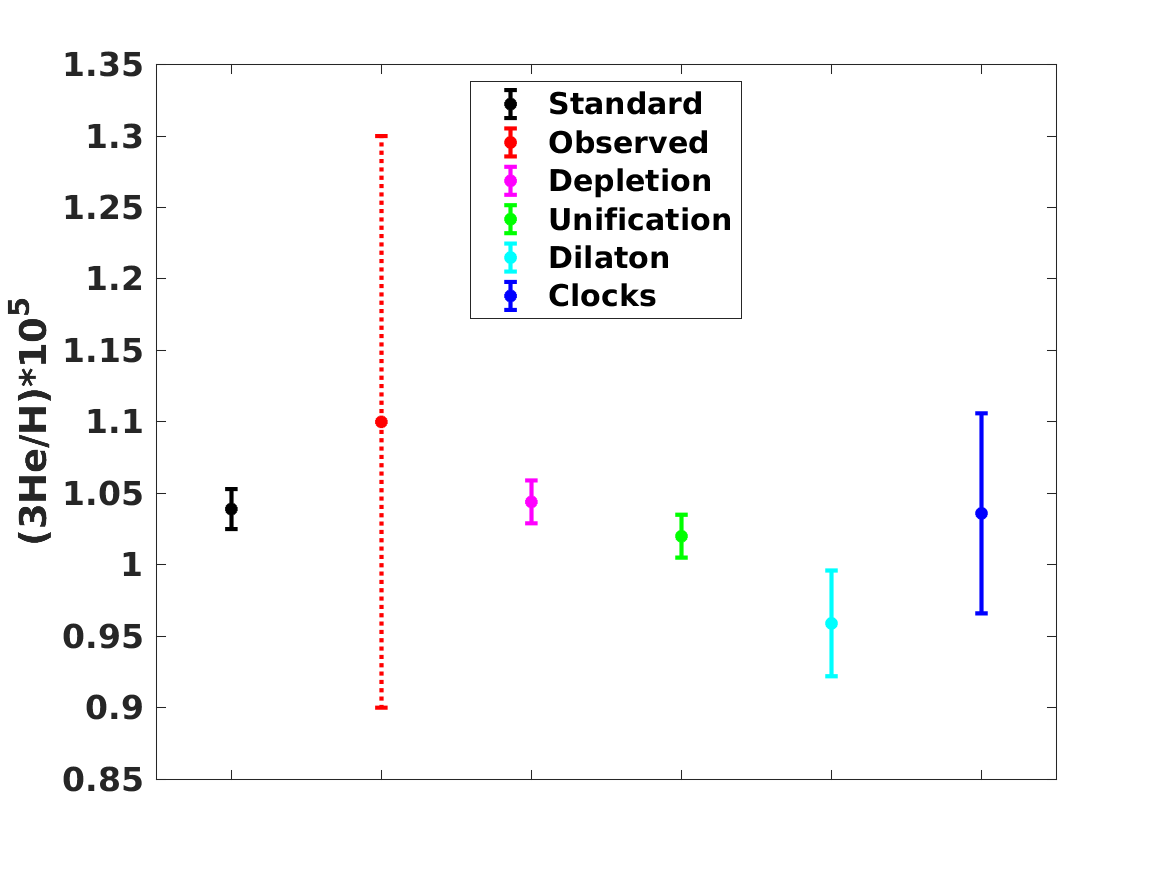}
\includegraphics[width=0.45\textwidth]{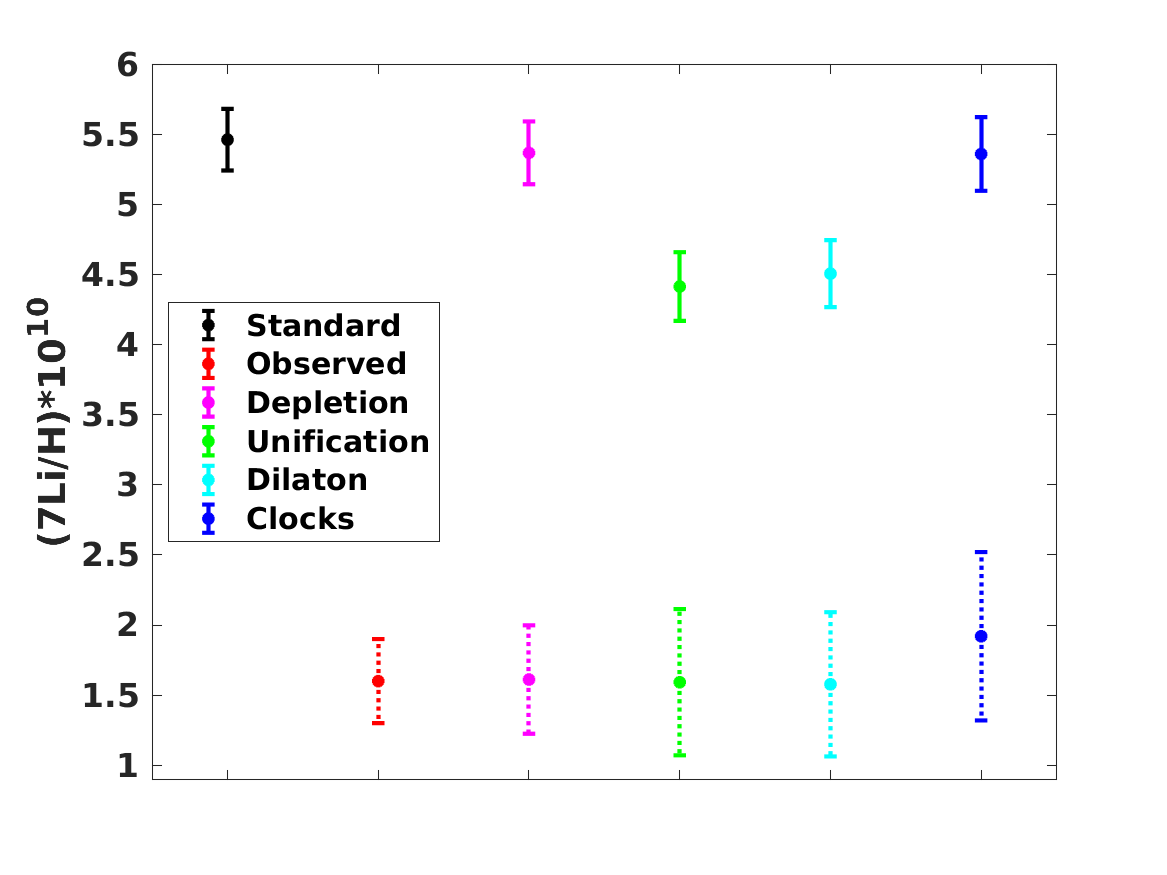}
\caption{Graphical comparison of the data in Tables \ref{table1} and \ref{table5} for each of the four nuclides. In each panel the black points are the standard theoretical values of \citet{Pitrou20}, while the red points are the observed abundances; both of these are listed in Table \ref{table1}. The magenta points show the best-fit depletion-only model discussed in Section \ref{cosmo}, while the green, cyan, and blue points are the best-fit in the Unification, Dilaton, and Clocks cases discussed in Table \ref{table5}. One sigma uncertainties have been depicted in all cases. The lithium-7 panel depicts both the cosmological and astrophysical (depleted) abundances; the latter, as well as the helium-3 observed abundance are shown with dotted error bars to indicate that they are not cosmological.}
\label{figure3}
\end{figure*}
%%%%%%%%%%%%%%%%%%%%

The observed BBN abundances allow the separate quantification of the role of both effects. We see that the preferred value of the depletion parameter decreases with respect to the one obtained with the standard value of $\alpha$ in Section \ref{cosmo}, by about one standard deviation. Moreover, there is still a preference for a positive value of $\Delta\alpha/\alpha$. Comparison with Table \ref{table4} shows that for the Unification and Dilaton models both the best-fit values and the statistical significance of the preference for the Baseline case are decreased with respect to the values found in Paper 1 and Paper 2, while for the Null case they are slightly increased. On the other hand, for the Clocks model the best-fit value for $\alpha$ is in agreement with the one found in Paper 1 and Paper 2 in both the Null and the Baseline cases. In this case the effect of the additional depletion mechanism is manifest in the best-fit values for the phenomenological particle physics parameters $R$ and $S$. Here we find
\be
R=15^{+15}_{-7}\,,\quad S=35^{+30}_{-16}\,,
\ee
whereas in Paper 2 the best-fit values were
\be
R=40^{+30}_{-10}\,,\quad S=100\pm30\,.
\ee
Note that although $R$ and $S$ are independent parameters in the likelihood analysis, the local prior from atomic clock experiments---cf. Equation (\ref{clopri})---effectively correlates them. The difference between the two sets of values reflects the fact that in Paper 2 there was no allowance for a depletion factor (effectively, $\Delta=0$ was used). Moreover, bearing in mind the aforementioned slight reduction of the depletion factor when a non-standard value of $\alpha$ is allowed, it also shows that although a value of $\Delta\alpha/\alpha>0$ does alleviate the lithium problem, it does not completely solve it. Instead, a significant depletion factor is still statistically preferred.

The more noteworthy aspect is that, unlike the results of Paper 1 and Paper 2, the preferred values for $\Delta\alpha/\alpha$ are almost identical in the Baseline and Null cases, meaning that the preference for a positive value of $\Delta\alpha/\alpha$ is not driven by the lithium-7 problem. In other words, even without including lithium-7 in the analysis, there is a mild statistical preference for a variation of $\Delta\alpha/\alpha$ at the few parts-per-million level. Instead, it is driven by the small discrepancy, already mentioned in the previous section, between the baryon-to-photon ratios inferred from BBN and the corresponding CMB value; in other words, it is driven by the helium-4 and deuterium data. We note that within the standard paradigm the baryon-to-photon ratio is a constant, having the same value at the BBN and CMB epochs; a different value at the two epochs necessarily implies the presence of new physics.

When adding lithium-7 to the analysis without allowing for the possibility of a depletion mechanism (as was done in Paper 1 and Paper 2) the preferred value of $\Delta\alpha/\alpha$ increases very significantly with respect to the Null case, since a larger value of $\alpha$ is needed solve the lithium problem per se. On the other hand, if a depletion mechanism is allowed (as in the present work) then this mechanism provides the dominant contribution to address the lithium problem, and the value of $\Delta\alpha/\alpha$ is very slightly changed with the respect to the null case. Nevertheless the inclusion of $\Delta\alpha/\alpha$ is still significant, which is manifest in the fact that the preferred depletion fraction changes with respect to the case discussed in the previous section by about one standard deviation.

A summary plot of the standard theoretical and observed abundances, together with those predicted for the best-fit values of the various models discussed both in Section \ref{cosmo} and in the present one, can be found in Figure \ref{figure3}.

It is interesting to note that although the Unification and Dilaton models lead to very similar lithium-7 abundances (a consequence of the fact that this nuclide drives the statistical analysis), they lead to significantly  different ones for the other nucildes, especially helium-4 and helium-3. Specifically, the Dilaton model leads to a significantly higher helium-4 abundance and a significantly lower helium-3 abundance, as compared to the Unification model. This is due to the fact that in the Dilaton model, where $R$ is comparatively large, there are large positive and negative sensitivity coefficients, respectively for helium-4 and helium-3. The comparison between the three models also confirms the well-known result that, for the observationally relevant ratios of the baryon-to-photon ratio, the deuterium and lithium-7 abundances are anticorrelated \citep{Anticorr1,Anticorr2}, so a reduction of the latter abundance leads to an increase of the former. Similarly, one can notice an anticorrelation between the helium-4 and helium-3 abundances, highlighting the fact that an accurate determination of the primordial cosmological abundance would provide a sensitive consistency test of BBN.

%%%%%%%%%%%%%%%%%%%%%%%%%%%%%%%%%%%%%%%%%%%%%%%%%%%%%%%%%%%%%%%
\section{Stellar depletion and the Lithium problem}
\label{depletion}

In this section we assess how realistic is the depletion factor $\Delta$, determined in the previous Section, according to stellar evolution theory. We compare $\Delta$ to predicted lithium-7 depletion in stellar models, including different transport processes of chemical elements.

\subsection{Input physics of the stellar models}

The depletion of lithium that occurs during the evolution of Population II stars is predicted using the Montreal/Montpellier stellar evolution code \citep{turcotte98} and Code d'Evolution Stellaire Adatpatif et Modulaire (CESTAM, the "T" stands for Transport: \citealt{morel08}, \citealt{marques13}, \citealt{deal18}). We used two stellar evolution codes firstly for confirmation of the results, and secondly because they do not include the same transport processes of chemical elements. The Montréal/Montpellier code has already been used to model Population II stars, and especially lithium surface abundances, including detailed atomic diffusion calculations \citep[e.g.][]{richard05}. The CESTAM evolution code also includes atomic diffusion (with independent formalisms, see \citealt{deal18} for more details) and is able to model the effect of rotation \citep{marques13}.

Our Montréal/Montpellier models have the same input physics as the ones used by \cite{deal21} without accretion. CESTAM models are computed with the same input physics as in \cite{deal20}, except for the mixing length parameter set to $\alpha_\mathrm{CGM}=0.69$ following the \citet{canuto96} formalism. The opacity tables are the OP table at fixed chemical composition and not the monochromatic ones \citep[OPCDv3.3,][]{seaton05}. We tested that with the additional transport processes included in the models, the abundance variation are not sufficient to modify the internal structure (especially the size of the surface convective zone). Using simpler opacity tables has then a negligible impact of the lithium surface abundance in the specific framework of this study\footnote{Such an approximation of the opacity calculation in presence of transport of chemical elements is not valid in most cases \citep[e.g.][]{turcotte98b,richard01,theado09,deal16,deal18}.}. No core overshoot is taken into account and the initial chemical composition is the one of Montréal/Montpellier models ([Fe/H]$_\mathrm{ini}=-2.31$~dex). 
Considering a wider range of initial chemical composition would not affect the conclusion of this work, because at these low metallicities the structure variations are small (especially the size of the surface convective zone, which is important for lithium surface abundances variations, see e.g. \citealt{richard02}). It should be noticed that, with such models, there is no extra depletion/dispersion for the lower metallicities, as it has been observed \citep{sbordone10}. The mechanism to explain this behavior is still unknown.
The models are computed with masses between 0.55 and 0.78~$M_\odot$. All models are on the main sequence at 12.5~Gyr and cover a range of effective temperature [5000,6500] K, typical of observed main-sequence Population II stars. The physics of all models is presented in Table~\ref{tab:inputs}.

%%%%%%%%%%%%%%%%%%%%%%%%%%%%%%%%%%%%%%%%%%%%%%%%%%%%%%%%%%%%%%
\begin{table*}
\caption{Input physics of the stellar models used in this work. All sets include models at 0.55, 0.60, 0.65, 0.70, 0.75 and 0.78~$M_\odot$. The complete initial chemical composition is available in the work of \cite{deal21}.}
\label{tab:inputs}
\centering
\begin{tabular}{l|c|c|c|c|c|c|c|c}
\hline\hline
Code &\multicolumn{1}{c}{Montr\'eal/Montpellier\tablefoottext{a}}&\multicolumn{7}{|c}{CESTAM\tablefoottext{b}}\\
\hline
Model set & 1 & 2 & 3 & 4 & 5 & 6 & 7 & 8  \\
\hline
$T_0$ [$10^6$~K]\tablefoottext{1} & \multicolumn{1}{c}{1.0, 1.32, 1.74, 1.80, 1.90, 2.24} & \multicolumn{1}{|c|}{1.60, 1.74, 1.80, 1.90} & - & - & - & - & - & - \\
$v_\mathrm{ZAMS}$ [km.$^{-1}$] & - & - & $15$ & $15$ & $15$ & $15$ & $15$ & $15$\\
$\nu_{v,\mathrm{add}}$ [cm$^2$.s$^{-1}$] & - & - & $10^8$ & $0$ & $10^8$ & $10^8$ & $10^8$ & $10^8$ \\
$\alpha_\mathrm{ovi}$ [$H_p$] & - & - & - & - & $0.1$ & $0.3$ & $0.4$ & $0.6$\\
\hline
Opacities & OPAL monochromatic\tablefoottext{c} &
\multicolumn{7}{|c}{OP\tablefoottext{d}}\\
Eos & CEFF\tablefoottext{e} &\multicolumn{7}{|c}{OPAL2005\tablefoottext{f}}\\
Nuc. react. & Bahcall92\tablefoottext{g} &\multicolumn{7}{|c}{NACRE+LUNA\tablefoottext{h}}\\
$\alpha_\mathrm{conv.}$ & 1.66\tablefoottext{i} &\multicolumn{7}{|c}{0.69\tablefoottext{j}}\\
\hline
Initial mix. & \multicolumn{8}{c}{AGSS09\tablefoottext{k}}\\
$X_\mathrm{ini}$ & \multicolumn{8}{c}{$0.7538$}\\
$Y_\mathrm{ini}$ & \multicolumn{8}{c}{$0.2461$}\\
($^7$Li/H)$\times 10^{10}$ & \multicolumn{8}{c}{$4.45$}\\

[Fe/H] & \multicolumn{8}{c}{$-2.31$}\\
$\alpha$ enhanc. & \multicolumn{8}{c}{$+0.3$}\\
Core ov. & \multicolumn{8}{c}{none}\\
\hline
\end{tabular}
\tablefoot{\tablefoottext{1}{Reference temperature in the expression of $D_\mathrm{turb}$ (see Equation~\ref{dturb}).}\tablefoottext{a}{\cite{turcotte98}.}\tablefoottext{b}{\cite{morel08,marques13,deal18}.}\tablefoottext{c}{\cite{iglesias96}, tables not public.}\tablefoottext{d}{\cite{seaton05}.}\tablefoottext{e}{\cite{JCD92}.}\tablefoottext{f}{\cite{rogers02}.}\tablefoottext{g}{\cite{bahcall92}.}\tablefoottext{h}{\cite{angulo99} + \cite{imbriani04} for $^{14}$N($p,\gamma$)$^{15}$O.}\tablefoottext{i}{Solar calibrated value with the \cite{bohm58} formalism.}\tablefoottext{j}{Solar calibrated value with the \cite{canuto96} formalism, for models including atomic diffusion.}\tablefoottext{k}{\cite{asplund09}.}}
\end{table*}
%%%%%%%%%%%%%%%%%%%%%%%%%%%%%%%%%%%%%%%%%%%%%%%%%%%%%%%%%%%%%%

\subsection{Transport processes of chemical elements}\label{transport}

We computed stellar models including different transport processes of chemical elements. In particular, they all affect the transport of lithium-7 and therefore play a role in the lithium depletion. Most of the processes considered in this study were extensively described by \citet{dumont20}, and references therein. In the following, we briefly describe the impact of the processes on lithium depletion and the prescriptions we use in the stellar models. As the aim of this section is to show that the $\Delta$ depletion factor can be explained by stellar physics, we only included simple modelling approaches for some of the processes addressed below. Including more realistic modelling is the next natural step of this study.\newline

\textbf{Convection:}

All models include convection using the Schwarzschild criterion. In convective zones, the transport of chemical elements is very efficient and fully homogenizes the chemical composition. If the bottom part of the convective zone has a temperature close to or larger than the temperature at which lithium is destroyed by proton capture ($\sim 2.5\times 10^6$~K), lithium is efficiently depleted. Convective boundaries are defined by the Schwarzschild criterion \citep{schwarzschild58} (or Ledoux criterion, \citealt{ledoux47}) in stellar models. It has been shown than convective plumes can penetrate the radiative zone and then extend the mixing of chemical elements beyond the standard convection criteria \citep[e.g.][]{zahn91,baraffe17}. If this extension of the mixing reaches the region where lithium is destroyed by proton capture, this will also lead to lithium depletion. For the models in this study, we include a simple constant step extension of the surface convective zone $\alpha_\mathrm{ovi}$ in pressure scale height units ($H_p$). More sophisticated approach exist \citep[see][for an overview]{dumont20}.\newline

\textbf{Atomic diffusion:}

All models also include atomic diffusion, which is a selective microscopic transport process. It occurs in every star and it the consequence of the internal gradients of pressure, temperature and concentration \citep{michaud15}. It is efficient in radiative zones. This is mainly the competition between gravity, which makes elements move toward the center of stars, and radiative acceleration, which makes elements move toward the surface of stars. For the stellar models used in this study, radiative acceleration on lithium is negligible. The transport of lithium by atomic diffusion is dominated by the gravitational settling.\newline

%%%%%%%%%%%%%%%%%%%%%%%%%%%%%%%%%%%%%%%%%%%%%%%%%%%%%%%%%%%%%%
\begin{figure*}
\centering
\includegraphics[width=0.49\textwidth]{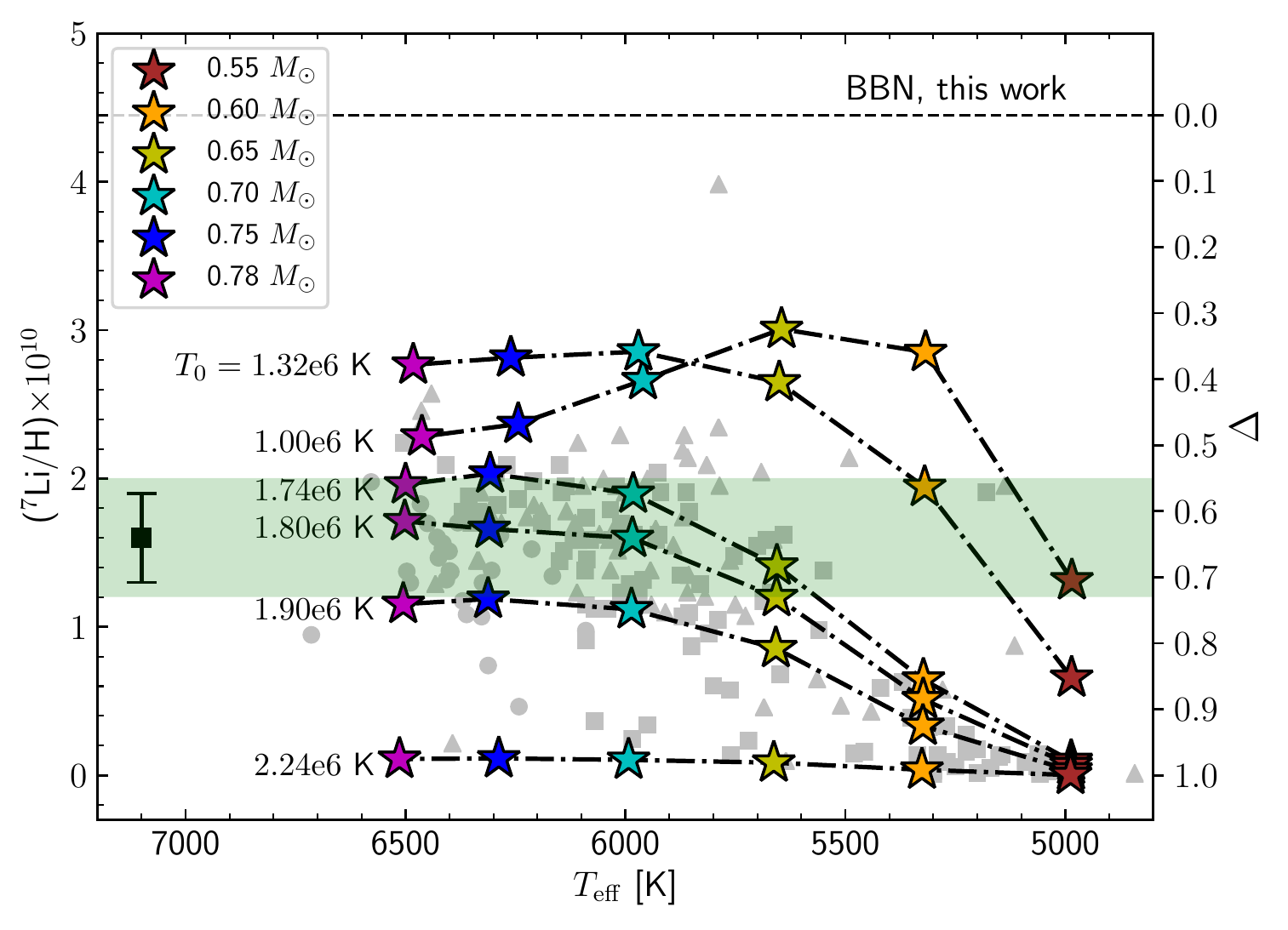}\hspace{0.25cm}\includegraphics[width=0.49\textwidth]{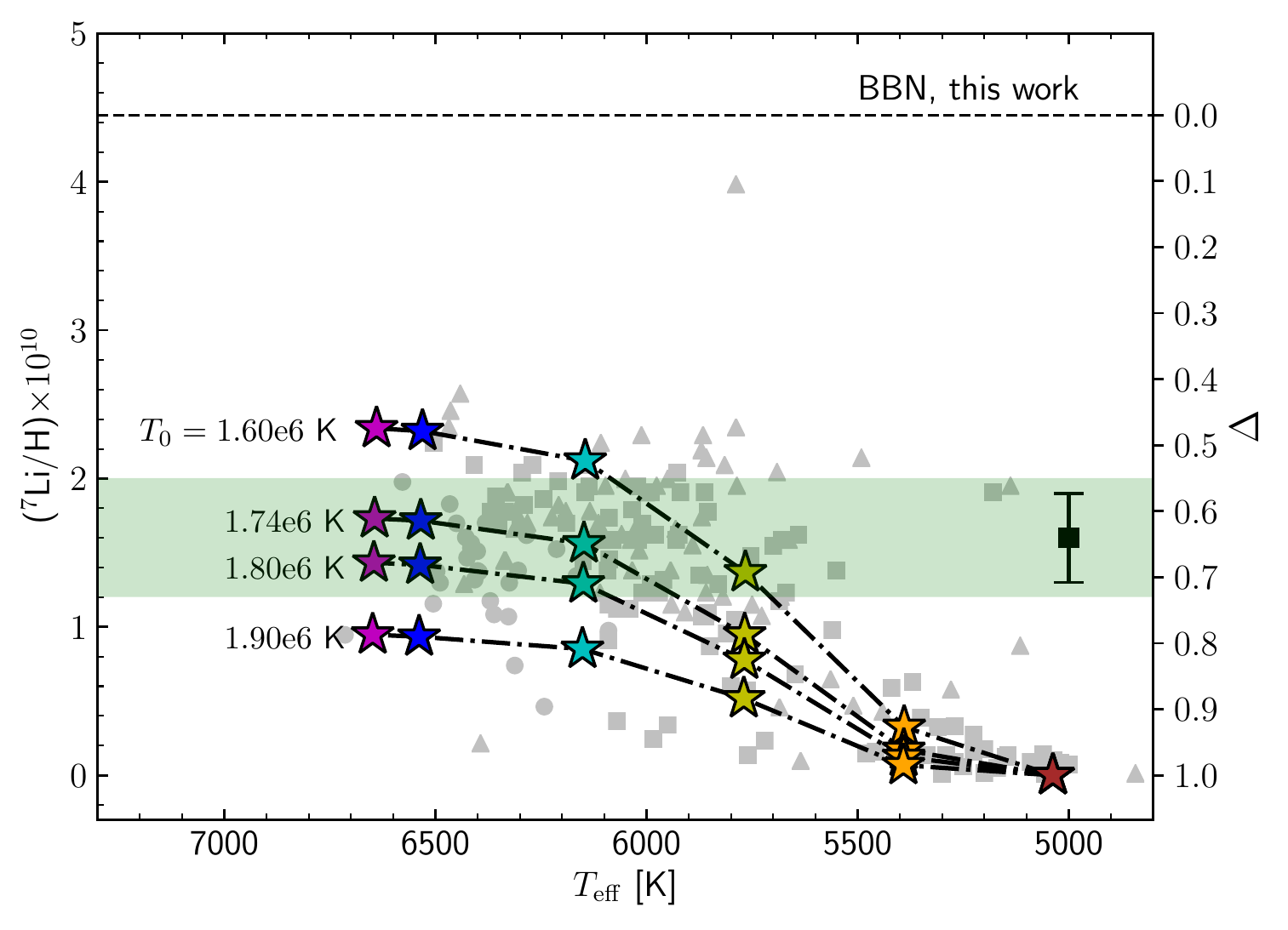}
\caption{Lithium abundance according to the effective temperature for Montr\'eal/Montpellier models (model set 1, left panel) and CESTAM models (model set 2, right panel) at 12.5~Gyr, taking into account different $T_0$ values for the parametrized turbulent diffusion coefficient. The corresponding $\Delta$ values are shown on the right y-axis. The grey symbols are observed lithium abundances from \cite{sbordone10} (circles), \cite{bonifacio97} (triangles) and the SAGA database \citep[http://sagadatabase.jp/,][]{suda08,suda11,yamada13,suda17} (squares). The black square with error bars represents the standard value for the lithium plateau \citep{sbordone10}. The green area represents the range of $\Delta$ values found in Section~\ref{constants}.}
\label{Li-MoMo-CESTAM}
\end{figure*}
%%%%%%%%%%%%%%%%%%%%%%%%%%%%%%%%%%%%%%%%%%%%%%%%%%%%%%%%%%%%%%

\textbf{Parametrized turbulent diffusion coefficient:}

Atomic diffusion alone cannot explain the lithium surface abundance of star. There is a need to include competing transport processes. It is possible to account for them using a parametrized turbulent diffusion coefficient. Such a coefficient was already used and calibrated to reproduce the lithium surface abundance in Population II stars \citep{richard05,deal21} and in clusters \citep[e.g.][]{gruyters13,gruyters16,korn07}. The turbulent diffusion coefficient is expressed as: 
\begin{equation}\label{dturb}
    D_\mathrm{turb}=400\times D_\mathrm{He}(T_0)~\left( \frac{\rho(T_0)}{\rho} \right)^3\,,
\end{equation}
where $T_0$ is a reference temperature. $D_\mathrm{He}(T_0)$ is the diffusion coefficient of helium at the reference point\footnote{The diffusion coefficient of helium can be easily calculated using this approximate formula: $D_{\rm He}=3.3 \times 10^{-15} T_0^{2.5} / [ 4 \rho \ln(1 + 1.125 \times 10^{-16} T_0^{3} /\rho)]$}, $\rho$ the local density and $\rho(T_0)$ the density at the reference point. The only free parameter of this parametrization is $T_0$. The way this parametrization is made, the transport of chemical elements is very efficient for internal temperatures smaller than $T_0$ (i.e. the chemical composition is homogenized from the surface down to $T_0$). For temperatures larger than $T_0$ the mixing decreases rapidly as a function of $\rho^{-3}$. The larger is $T_0$, the deeper goes the efficient mixing region.\newline

\textbf{Rotational-induced mixing:}\newline

The transport of chemical elements induced by rotation is one of the processes in competition with atomic diffusion. It has an important impact on chemical elements and especially on lithium \citep[e.g.][]{charbonnel99,talon08,dumont20}. Rotational-induced mixing is mainly driven by the shear instability and the meridional circulation. The differences between the models of this study and the ones of \cite{deal20} are the horizontal diffusion coefficient ($D_h$, from \citealt{mathis18}) and an additional vertical viscosity $\nu_v=10^8$~cm$^2$s$^{-1}$ as calibrated by \citet{ouazzani19} to take into account the fact that the current rotation theory underestimates the transport of angular momentum. We also tested smaller values of $\nu_v$. All models including rotation have a rotation speed at the ZAMS $v_\mathrm{ZAMS}=15$~km.$^{-1}$. See \cite{marques13} for a detailed description of the modelling of rotation in CESTAM.

\subsection{Parametrized turbulent transport and atomic diffusion}

We first estimate the required transport of chemical elements to explain the $\Delta$ depletion factor, using the parametric turbulent diffusion described in Section~\ref{transport} and atomic diffusion. This kind of approach has already been used to study lithium surface abundances of Population II stars \citep[e.g.][]{richard05,deal21}.

The left panel of Fig.~\ref{Li-MoMo-CESTAM}, shows the lithium surface abundance at 12.5~Gyr in stellar models computed with the Montr\'eal/Montpellier stellar evolution code, including atomic diffusion and different parametrizations of the turbulent diffusion coefficient (models set 1). The models are computed with an initial primordial lithium abundance ($^7$Li/H)$\times 10^{10}$= 4.45, following the values of Table~\ref{table5} for the Unification and Dilaton models. Lithium depletion is larger for the smaller masses because of a deeper surface convective zone, i.e. close to the region where lithium is destroyed by nuclear reactions. The larger the $T_0$ value, the deeper the turbulent mixing, leading to stronger depletion. The smaller the $T_0$ value, the stronger are the effect of atomic diffusion, i.e. the turbulent mixing is not strong enough to balance it. This is the reason why for $T_0=1.0\times10^6$~K, the depletion is larger at 0.78 than 0.65~$M_\odot$. We see that observations, as well as the range of $\Delta$ values previously determined, are well reproduced by the models, for the whole effective temperature range, for $T_0$ values between $1.74\times 10^6$ and $1.90\times 10^6$~K.

The CESTAM models (right panel of Fig.~\ref{Li-MoMo-CESTAM}, models set 2), show similar surface abundances, with a slight shift in the $T_0$ values and in effective temperatures. This difference comes from the different input physics (including the equation of state, opacity tables, and nuclear reaction rates) between the two stellar evolution codes. At a given $T_0$ value, the difference in lithium abundance is about 10\% at maximum, which will not impact the conclusion of this study. This justifies our confidence in using CESTAM models. In the following subsections, we will estimate what are the possible physical processes responsible for such turbulent transport.

%%%%%%%%%%%%%%%%%%%%%%%%%%%%%%%%%%%%%%%%%%%%%%%%%%%%%%%%%%%%%%%%%%%%%%%%%%%%%%

\subsection{Rotationally-induced mixing and atomic diffusion}

%%%%%%%%%%%%%%%%%%%%%%%%%%%%%%%%%%%%%%%%%%%%%%%%%%%%%%%%%%%%%%
\begin{figure}
\centering
\includegraphics[width=0.49\textwidth]{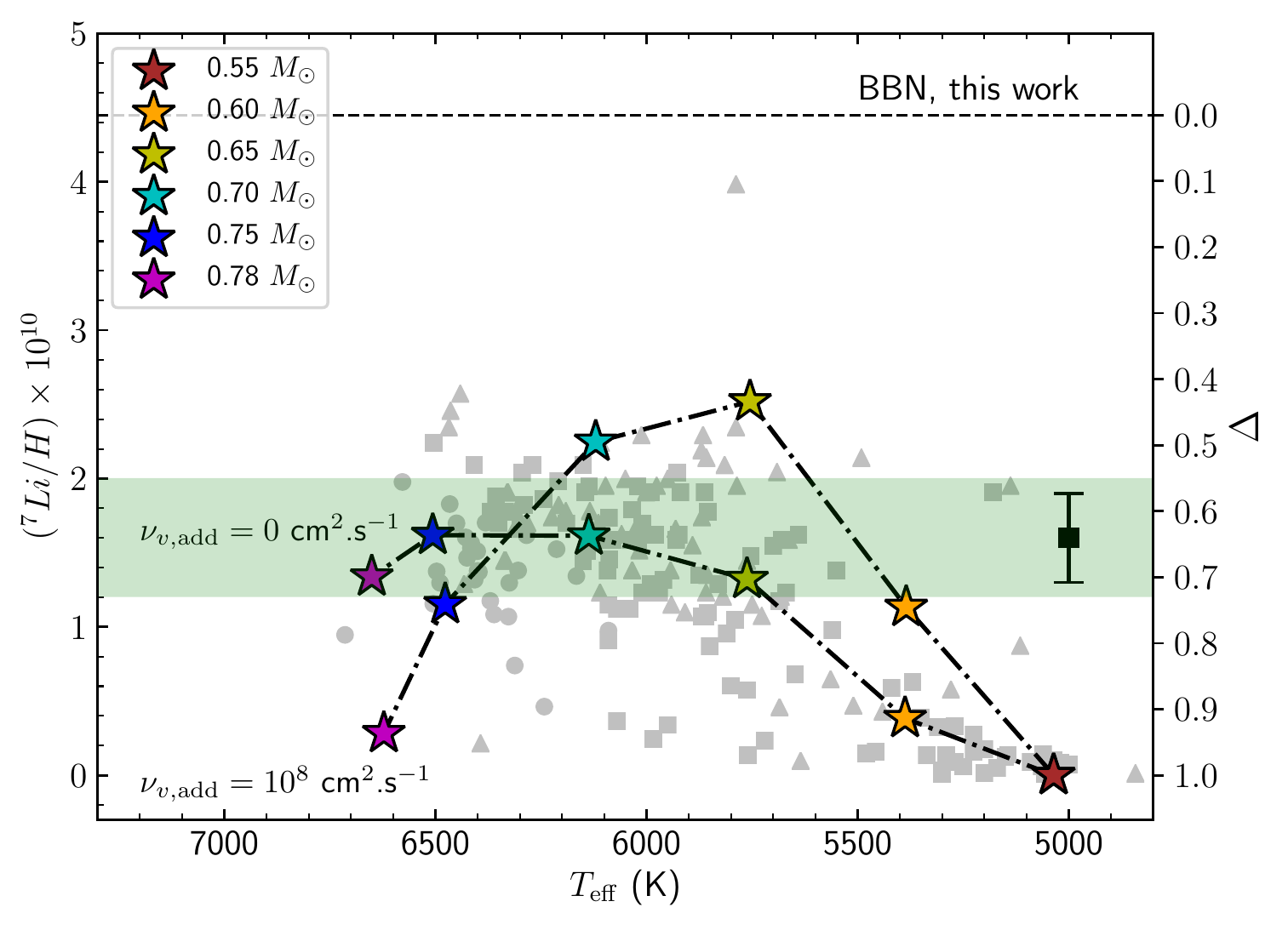}
\caption{Lithium abundance according to the effective temperature for stellar models of sets 3 and 4 at 12.5~Gyr, taking into account different $\nu_{v,\mathrm{add}}$ values. Same legend as Fig.~\ref{Li-MoMo-CESTAM} for the other symbols and the green area.}
\label{Li-rota}
\end{figure}
%%%%%%%%%%%%%%%%%%%%%%%%%%%%%%%%%%%%%%%%%%%%%%%%%%%%%%%%%%%%%%

In this section we assess whether the $\Delta$ depletion factor is linked to the combination of rotational induced-mixing and atomic diffusion. Figure~\ref{Li-rota} shows two sets of models (sets 3 and 4) in which the two processes are included. In one case, rotation is almost solid (set 3, with $\nu_{v,\mathrm{add}}=10^8$~cm$^2$s$^{-1}$) and in the other case the internal rotation is differential using the standard theory for rotation (set 4, with $\nu_{v,\mathrm{add}}=0$~cm$^2$s$^{-1}$). The models with an internal differential rotation are in very good agreement with the observations. We tested larger initial rotation speeds for the models with quasi-solid rotation and the results are very similar. The rotation speed does not play a major role for the transport of chemicals in this specific case.

Helio/Asteroseismology (study of oscillation of the Sun/stars) also allows us to probe the internal rotation of stars. It has been shown that the internal rotation of the Sun is nearly uniform in the radiative zone at least down to $R=0.2~R_\odot$ \citep[e.g.][]{kosovichev88,garcia07}. Solar models accounting for standard rotation predict a high degree of radial differential rotation \citep[e.g.][]{eggenberger05}. The same conclusions are reached when comparing other stars for which we can access the core rotation \citep[e.g.][]{tayar13,deheuvels14,ouazzani19}. This indicates that among sets 3 and 4, the more realistic models are those of set 3. This implies that the lithium surface abundances of Population II cannot be explained by the combined effect of atomic diffusion and rotation. The same kind of conclusion have been previously obtained for other types of stars \citep[e.g.][]{deal20}.

\subsection{Penetrative convection, rotationally-induced mixing and atomic diffusion}\label{diff-rota-ovi}

%%%%%%%%%%%%%%%%%%%%%%%%%%%%%%%%%%%%%%%%%%%%%%%%%%%%%%%%%%%%%%
\begin{figure}
\centering
\includegraphics[width=0.49\textwidth]{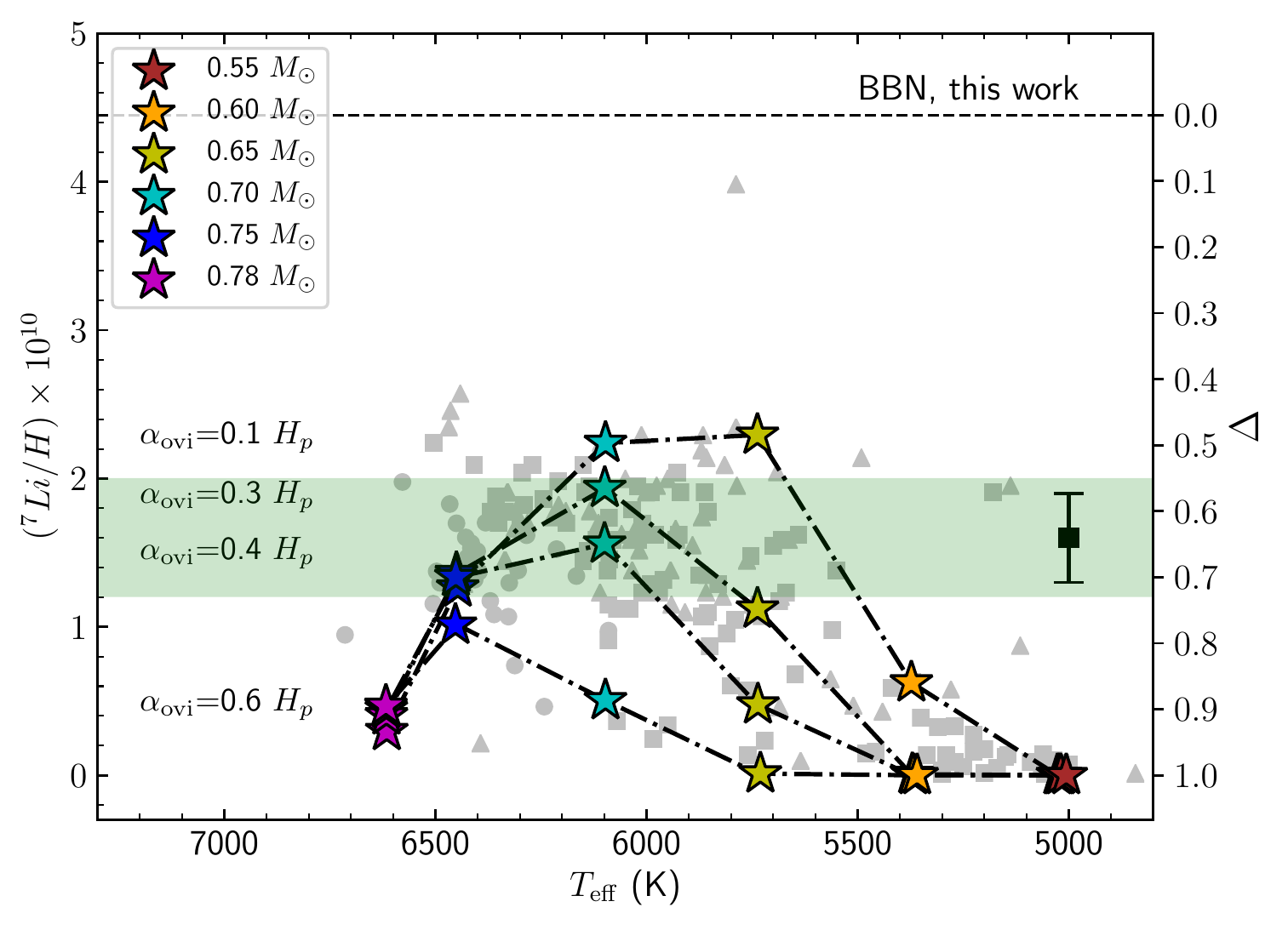}
\caption{Lithium abundance according to the effective temperature for stellar models of sets 5, 6, 7 and 8, all at 12.5~Gyr, taking into account different $\alpha_\mathrm{ovi}$ values. Same legend as Fig.~\ref{Li-MoMo-CESTAM} for the other symbols and the green area.}
\label{Li-rota-ovi}
\end{figure}
%%%%%%%%%%%%%%%%%%%%%%%%%%%%%%%%%%%%%%%%%%%%%%%%%%%%%%%%%%%%%%

In this section we assess whether the $\Delta$ depletion factor can be accounted for by the combination of penetrative convection, rotational induced-mixing and atomic diffusion. Figure~\ref{Li-rota-ovi} shows four sets of models (set 5, 6, 7 and 8) in which the three processes are included. We see that the four sets cover well the range of observed lithium surface abundances up to an effective temperature of about $6000$~K. The values of penetrative convection considered in the models are consistent with what has been determined for the Sun ($\alpha_\mathrm{ovi}\sim 0.37~H_p$, \citealt{christensen11}). For the hotter (more massive) stars, an additional transport process is probably needed to reduce the lithium depletion induced by atomic diffusion. It is possible that the amount of penetrative convection is larger for the more massive stars, which could also make the models agree better with the observation. A more realistic treatment of this process could also improve the agreement. Note that penetrative convection was also invoked in an other scenario to explain lithium abundances in Population II stars \citep[e.g.][]{fu15}.

\subsection{A minimum astrophysical depletion?}

It is interesting to note that there is a minimum depletion obtained in the models presented in Section~\ref{diff-rota-ovi}, which is around $\Delta_\mathrm{min}=[0.4;0.5]$. From the stellar modelling point of view, the transport process with the more accurate modelling and leading to the smallest lithium depletion (for masses around 0.60-0.70~M$_\odot$) is atomic diffusion. The efficiency of atomic diffusion strongly depends on the internal structure of stars, and especially on the size of the surface convective zone. The prediction of the minimum depletion of lithium, expected from stellar models including atomic diffusion only, is then subject to the input physics of the models (including opacity tables, equations of state, convection theory, nuclear reaction rates, etc.). This is less the case when other competing transport processes are taken into account (see the comparison done in Fig.~\ref{Li-MoMo-CESTAM}). In this context, it is difficult to provide a strong constrain on the minimum lithium depletion expected from stellar models. As an example, for the input physics considered in this paper, we find $\Delta_\mathrm{min}=[0.25;0.40]$ for models including atomic diffusion only. The only robust conclusion we can draw at this point is that $\Delta_\mathrm{min}>0$ in all models including expected transport processes (i.e. at least atomic diffusion).

Overall, we have shown that lithium surface abundances of Population II stars, as well as the range of $\Delta$ values determined in Section~\ref{constants}, are well reproduced by models including atomic diffusion, a nearly uniform rotation, and some amount of penetrative convection convection, even if our modelling of the latter and of the missing transport of angular momentum has been somewhat simple. A more realistic modelling of these processes, and a systematic study of the effect of the different input physics (opacity tables, equation of states, convection theory, nuclear reaction rates, etc.), are needed to drawn stronger conclusions, but our work suffices to show that transport processes of chemical elements in stars not only need to be considered when addressing the lithium problem, but also are very probable candidates to solve it.

%%%%%%%%%%%%%%%%%%%%%%%%%%%%%%%%%%%%%%%%%%%%%%%%%%%%%%%%%%%%%%

\section{Varying fundamental constants and the Deuterium discrepancy}
\label{deuterium}

Having previously identified the small discrepancy between the baryon-to-photon ratios inferred from BBN and the CMB value as the origin of the preference for a non-standard value of $\alpha$, here we assess the relative contributions of the helium-4 and deuterium data to this result.

%%%%%%%%%%%%%%%%%%%%
\begin{figure*}
\centering
\includegraphics[width=0.45\textwidth]{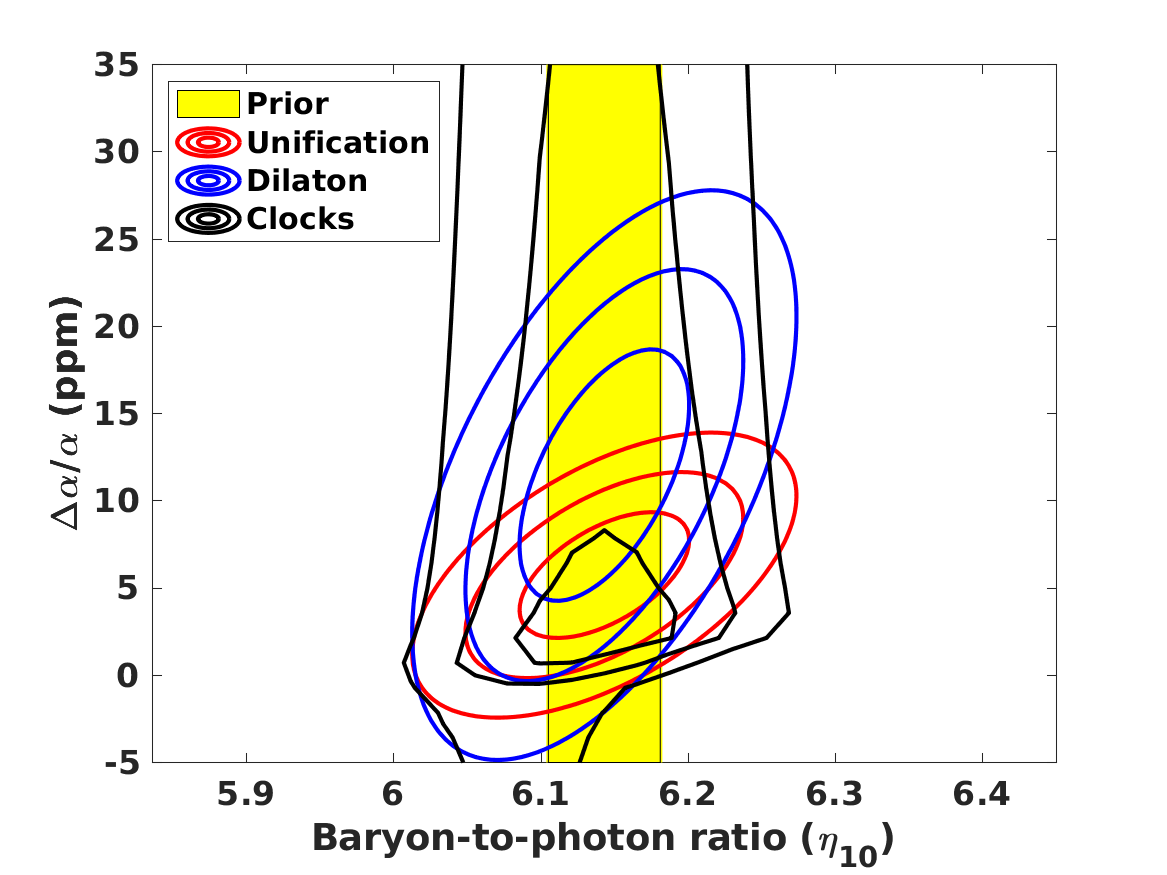}
\includegraphics[width=0.45\textwidth]{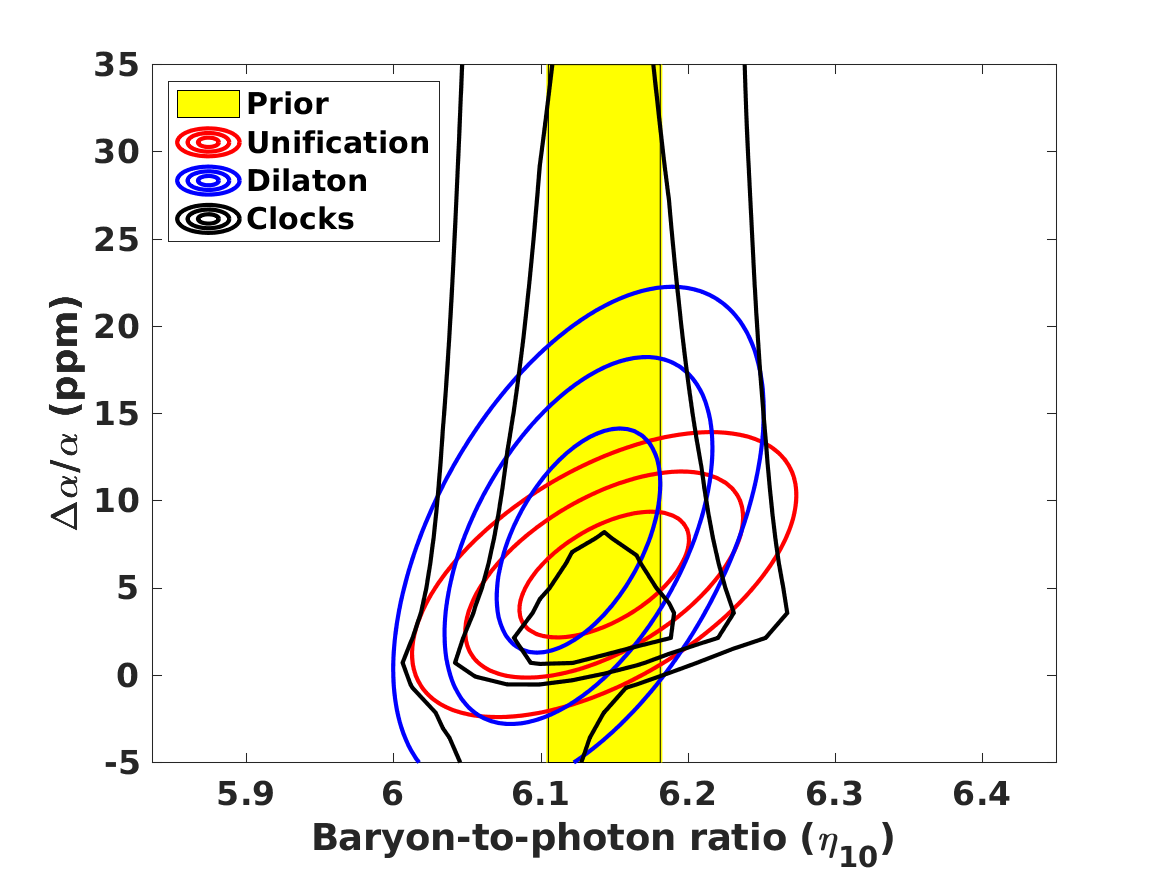}
\includegraphics[width=0.45\textwidth]{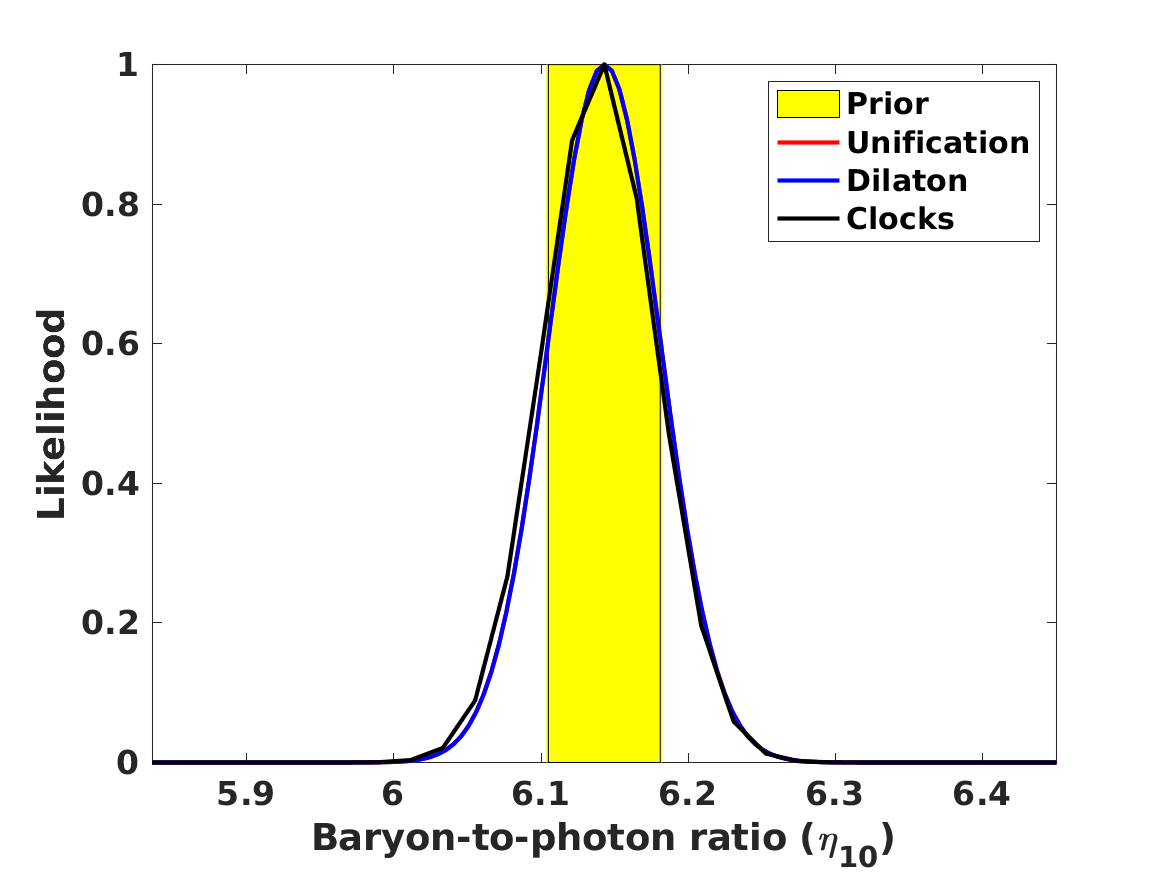}
\includegraphics[width=0.45\textwidth]{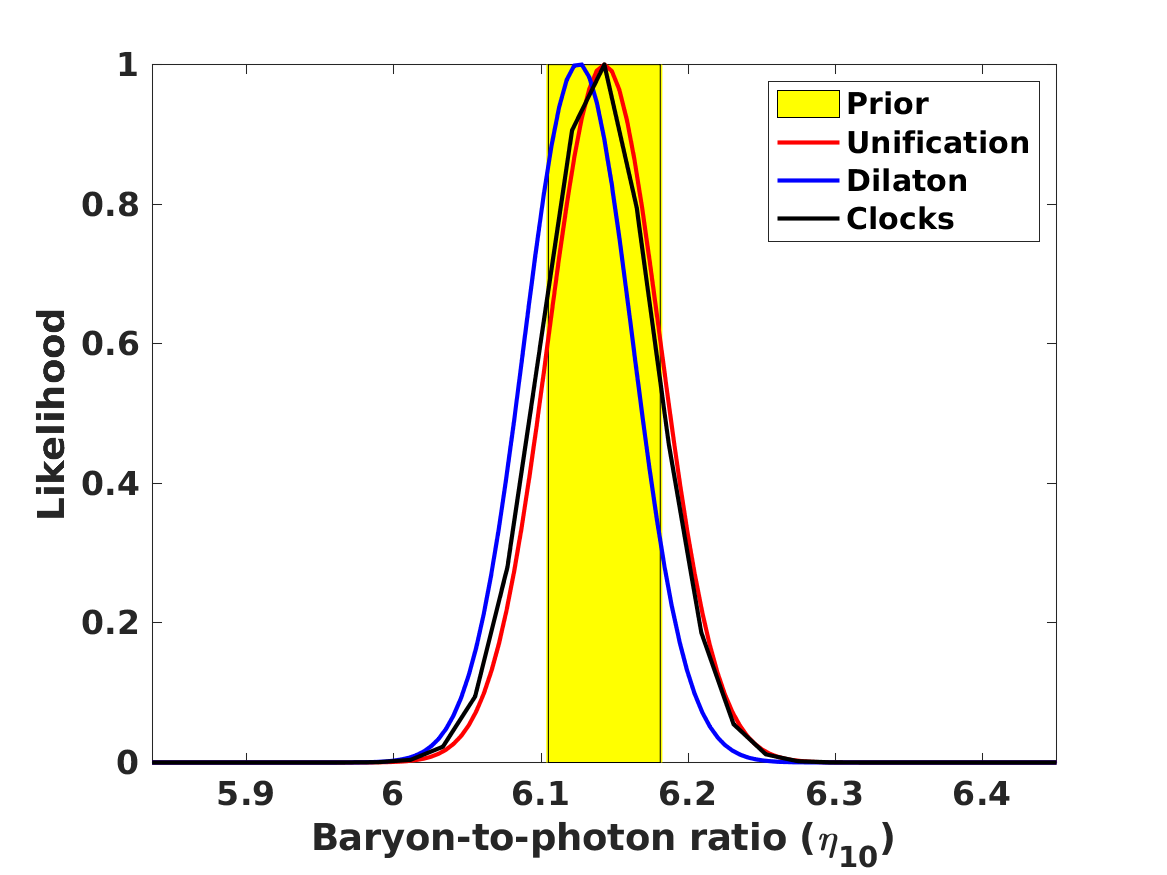}
\includegraphics[width=0.45\textwidth]{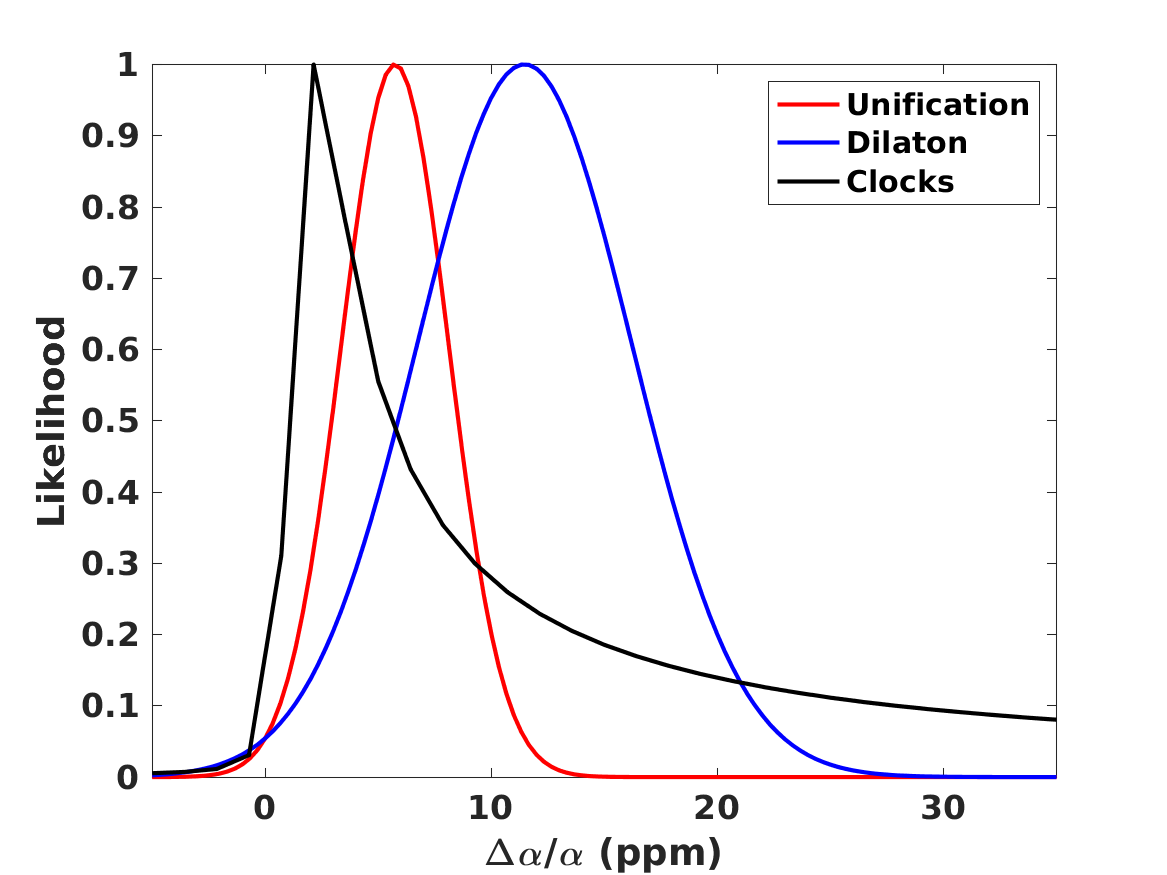}
\includegraphics[width=0.45\textwidth]{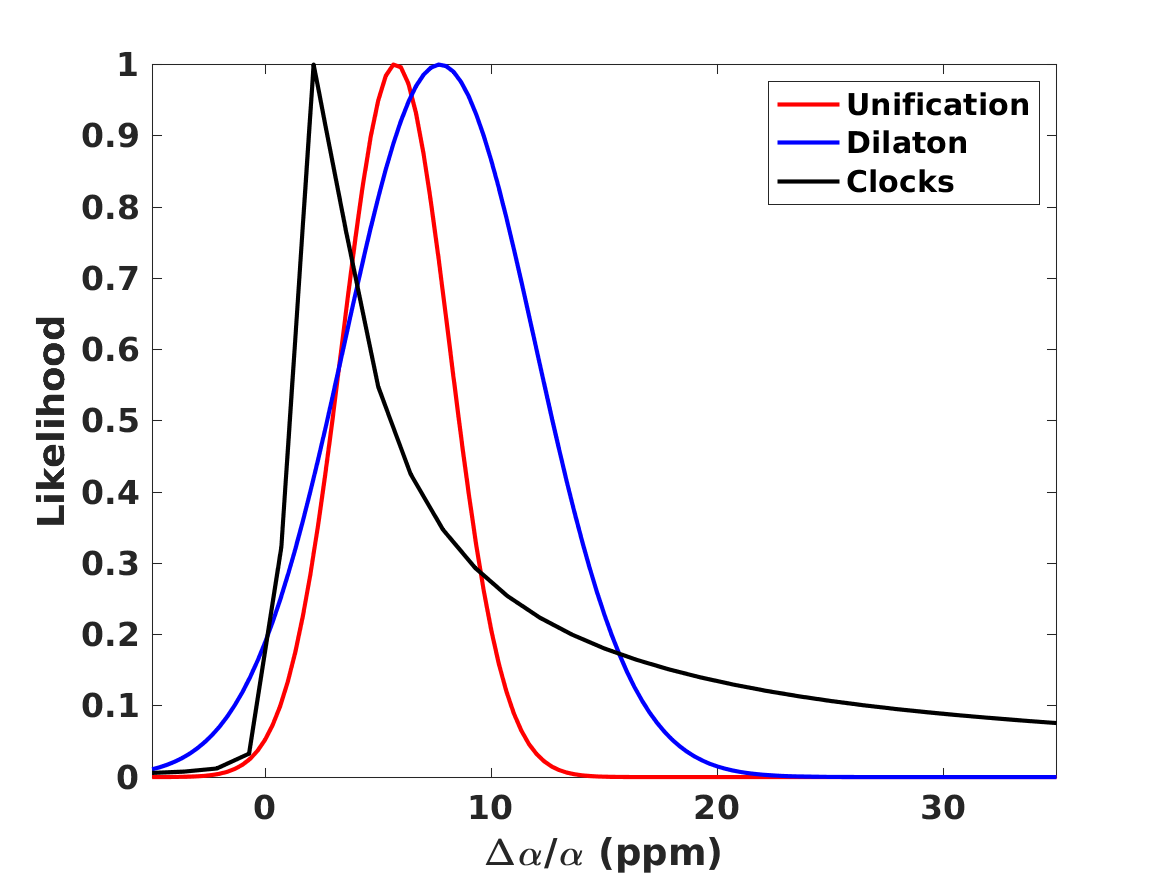}
\caption{BBN constraints on the fine-structure constant and the baryon-to-photon ratio, with remaining parameters marginalized in each case. The two-dimensional planes depict the 68.3, 95.4 and 99.7 percent confidence levels. Left side panels show the results for Deuterium only, and right side panels show the results for Deuterium plus Helium4.}
\label{figure9}
\end{figure*}
%%%%%%%%%%%%%%%%%%%%

To this end, we repeat the analysis both for the case where only the deuterium abundance is used and for the case where both deuterium and helium-4 are used. For each of these we further consider two sub-cases, ether keeping the cosmological parameters (i.e., the neutron lifetime, number of neutrinos and baryon-to-photon ratio) fixed at their best-fit values introduced in Sect. \ref{cosmo}, or letting these parameters vary (with the previously described priors) and then marginalizing them.

In the former case, the parameter space is summarized by Eq. \ref{paramalpha} and the relevant sensitivity coefficients are given in Table \ref{table3}. In the latter case the parameter space is wider:
\begin{equation}\label{fullparam}
\frac{\Delta Y_i}{Y_i}=(x_{n,i}+y_{n,i}S+z_{n,i}R)\frac{\Delta\alpha}{\alpha}+t_i\frac{\Delta\tau_n}{\tau_n}+v_i\frac{\Delta N_\nu}{N_\nu}+w_i\frac{\Delta\eta}{\eta}\,.
\end{equation}
As previously mentioned, a variation of $\alpha$ itself impacts the neutron lifetime; unlike in Sect. \ref{constants}, here we explicitly separate this effect from the rest of the effects of the $\alpha$ variation. This means that the $\alpha$-related sensitivity coefficients will change, and we have highlighted this point by adding the $n$ subscript in the previously defined sensitivity coefficients $x$, $y$ and $z$. For completeness, all these sensitivity coefficients are reproduced in Table \ref{table8}.

%%%%%%%%%%%%%%%%%%%%%%%%%%%%%%%%%%%%%%%%%%%%%%%%%%%%%%%%%%%%%%%%%%%%%%%%%%%%%%
\begin{table}
\caption{Sensitivity coefficients of BBN nuclide abundances on the free parameters of our full phenomenological parametrisation, defined in Eq. \ref{fullparam} the main text.}
\label{table8}
\centering
\begin{tabular}{| c | c c c c |}
\hline
$C_{ij}$ & D & ${}^3$He & ${}^4$He & ${}^7$Li \\
\hline
$x_{n,i}$ & +42.1 & +1.30 & -4.5 & -166.5 \\
$y_{n,i}$ & +40.1 & +2.00 & -3.5 & -150.7 \\
$z_{n,i}$ & +34.9 & -90.0 & +11.8 & -202.6 \\
\hline
$t_i$ & +0.442 & +0.141 & +0.732 & +0.438 \\
$v_i$ & +0.409 & +0.136 & +0.164 & -0.277 \\
$w_i$ & -1.65 & -0.567 & +0.039 & +2.08 \\
\hline
\end{tabular}
\end{table}
%%%%%%%%%%%%%%%%%%%%%%%%%%%%%%%%%%%%%%%%%%%%%%%%%%%%%%%%%%%%%%%%%%%%%%%%%%%%%%

The results of our analysis with the deuterium and helium-4 abundances are summarized in Figure \ref{figure9} and Table \ref{table9}; the latter can be usefully compared with the Baseline and Null cases reported in Table \ref{table5}.

%%%%%%%%%%%%%%%%%%%%%%%%%%%%%%%%%%%%%%%%%%%%%%%%%%%%%%%%%%%%%%%%%%%%%%%%%%%%%%
\begin{table}
\caption{Constraints on $\Delta\alpha/\alpha$ for the Unification, Dilaton and Clocks models (with the range of values within $\Delta\chi^2=1$ of it, corresponding to the $68.3\%$ confidence level for a Gaussian posterior likelihood), obtained from the Deuterium only and Deuterium plus Helium4 abundances, and with the values of the other relevant parameters (neutron lifetime, number of neutrinos and baryon-to-photon ratio) fixed or marginalized as described in the main text.}
\label{table9}
\centering
\begin{tabular}{| c | c | c | c | c |}
\hline
Data & Parameters & Unification & Dilaton & Clocks \\
\hline
D only & Fixed & $5.8\pm2.0$ & $10.9\pm3.8$ & $2.3^{+2.1}_{-0.8}$\\
D only & Marginalized & $5.7\pm2.7$ & $11.3\pm4.8$ & $2.1^{+2.9}_{-0.9}$\\
\hline
D $+$ He4 & Fixed & $5.8\pm2.0$ & $7.7\pm3.4$ & $2.2^{+2.4}_{-0.8}$\\
D $+$ He4 &Marginalized & $5.7\pm2.7$ & $7.7\pm4.3$ & $2.1^{+2.7}_{-0.9}$\\
\hline
\end{tabular}
\end{table}
%%%%%%%%%%%%%%%%%%%%%%%%%%%%%%%%%%%%%%%%%%%%%%%%%%%%%%%%%%%%%%%%%%%%%%%%%%%%%%

First and foremost, we confirm that the preference for $\Delta\alpha/\alpha>0$ is driven by the Deuterium results, and the top panels of Figure \ref{figure9} also makes it clear that this is due to the positive correlation between $\Delta\alpha/\alpha$ and the baryon-to-photon ratio $\eta$, though the strength of this correlation is model-dependent. Secondly, we find that the Helium-4 abundance plays a relatively minor role, though one that again depends on the assumed model.

Specifically, for the Unification and Clocks model, the results in Table \ref{table9} are fully consistent with the ones for the previously discussed Baseline and Null cases. The positive correlation between $\Delta\alpha/\alpha$ and $\eta$ is relatively mild, In these circumstances helium-4 does not noticeably affect the constraints on $\Delta\alpha/\alpha$, while allowing the cosmological parameters to vary and marginalizing them simply increases the statistical uncertainties, and therefore decreases the significance of the $\Delta\alpha/\alpha>0$ preference.

On the other hand, for the Dilaton model the results are more sensitive to the assumptions underlying the analysis. On the one hand, the stronger positive correlation between $\Delta\alpha/\alpha$ and $\eta$ means that whether or not $\eta$ is allowed to vary has a mild but discernible impact even in the case where only deuterium is used in the analysis. On the other hand, the inclusion of helium-4 significantly lowers the preferred value of $\Delta\alpha/\alpha$, the reason being that this model tends to overpredict the helium-4 abundance, as can bee seen in Figure \ref{figure3}. Again, this shows the potential of BBN as a precision test of these theoretical scenarios.

%%%%%%%%%%%%%%%%%%%%%%%%%%%%%%%%%%%%%%%%%%%%%%%%%%%%%%%%%%%%%%%
\section{Conclusions}
\label{concl}

We have provided an updated analysis of BBN constraints in the framework of a wide class of Grand Unified Theory scenarios \citep{Clara,Martins}, in particular addressing the long-standing Lithium problem \citep{Fields,Mathews} and the more recently noticed Deuterium discrepancy \citep{Pitrou20}. For the former we have highlighted and quantified the astrophysical mechanisms that can provide a solution, while for the latter we have identified a possible hint of new physics, specifically a mild (two to three standard deviations) preference for $\Delta\alpha/\alpha>0$, at the parts per million level. Such variations would be consistent with all other current $\alpha$ constraints \citep{ROPP}.

We note that the Deuterium discrepancy identified by \citet{Pitrou20} is not found by other authors \citep{Yeh,Pisanti}, who rely on different reaction rates. From a statistical point of view, the main difference is that the uncertainties in the theoretical abundances obtained by \citet{Pitrou20} are smaller than those of other analyses, although there are also small differences in the preferred values of these abundances. While we have not repeated our analysis with these different reaction rates (and the corresponding theoretical abundances inferred from them) we would expect that the propagation of the larger uncertainties on the theoretical side would lead to larger uncertainties on the fine-structure constant measurements, reducing the statistical significance of the preference for a non-standard value.

We have quantitatively determined the amount a lithium depletion needed to solve the lithium problem and searched for depletion mechanisms occurring during the evolution of stars, namely, the transport processes of chemical elements. We have shown that taking into account, in stellar models, atomic diffusion, rotational-induced mixing and some amount of penetrative convection (consistent with the amount determined from solar observation, \citealt{christensen11}), allows us to reproduce the lithium surface abundances of Population II stars. More realistic modelling of some of the processes (especially the penetrative convection, and the missing transport of angular momentum) are required to draw a more robust conclusion, but this work shows the necessity of including the stellar contribution to the lithium problem. Transport processes of chemical elements are most likely the dominant contribution to the solution to this problem.

The stellar models were computed with the best fit value for the initial primordial lithium abundance derived in this study, in the context of the GUT scenarios under consideration, of approximately ($^7$Li/H)$\times10^{10}=4.45$. Considering stellar models with a larger initial primordial lithium, as obtained in the standard BBN model (i.e. ($^7$Li/H)$\times10^{10}=5.46$) would not strongly impact our conclusions. This difference would only lead to the need of a slightly larger amount of penetrative convection to explain the extra depletion. This also shows that from the stellar physics point of view, the accuracy of the primordial lithium abundance is crucial to constrain the transport of chemical elements in metal poor stars.

It is worthy of note that currently there are only two primordial (cosmological) abundances that are well known, those of Deuterium and Helium-4. It is therefore highly desirable to close the loop through a cosmological measurement of the Helium-3 abundance, especially given the anticorrelation between the helium-4 and helium-3 abundances. This would enable a key consistency test of the underlying physics, which will be particularly crucial should evidence for new physics be confirmed.

Our analysis highlights the role of BBN as consistency test of the standard cosmological paradigm, and as a sensitive probe of new physics. For the broad phenomenological but physically motivated class of models we have considered, improving the observed abundances of Deuterium and Helium-4 by a factor of 2-3 will provide a stringent test, and in particular will definitively confirm or rule out the present tentative evidence for the $\alpha$ variation. This is a highly compelling science case for the forthcoming Extremely Large Telescope \citep{HIRES,SPIE}, provided it has an efficient blue wavelength coverage.

%%%%%%%%%%%%%%%%%%%%%%%%%%%%%%%%%%%%%%%%%%%%%%%%%%%%%%%%%%%

\begin{acknowledgements}
This work was supported by FCT---Funda\c c\~ao para a Ci\^encia e a Tecnologia through national funds (grants PTDC/FIS-AST/28987/2017 and PTDC/FIS-AST/30389/2017) and by FEDER---Fundo Europeu de Desenvolvimento Regional funds through the COMPETE 2020---Operacional Programme for Competitiveness and Internationalisation (POCI-01-0145-FEDER-028987 and POCI-01-0145-FEDER-030389). Additional funds were provided by FCT/MCTES through the research grants UIDB/04434/2020 and UIDP/04434/2020. MD is supported by national funds through FCT in the form of a work contract. CJM gratefully acknowledges useful discussions with Brian Fields and Cyril Pitrou in the context of the ESO Cosmic Duologue on BBN, and with Paolo Molaro. We thank an anonymous referee for valuable comments which helped to improve the paper.

\end{acknowledgements}

\bibliographystyle{aa} % style aa.bst
\bibliography{bbn} % your references Yourfile.bib
\end{document}